\documentclass[5p]{elsarticle}
\usepackage{newtxtext}  
\usepackage{amsmath}
\usepackage{amssymb}
\usepackage{graphicx}
\usepackage{subcaption}
\graphicspath{{scripts/0-schematics/}{scripts/1-first-order-interface/}{scripts/2-higher-order-reactions/}{scripts/3-competing-channels/}{scripts/4-hyperion/}}
\usepackage{xcolor}
\usepackage{todonotes}
\usepackage{overpic}
\usepackage{wasysym}
\usepackage[dvipsnames]{xcolor}
\usepackage{enumitem}
\usepackage[colorlinks = true, linkcolor = blue, urlcolor  = blue, citecolor = blue]{hyperref}
\usepackage{pythonhighlight}

\setlist[itemize]{noitemsep, topsep=3pt, parsep=2pt, leftmargin=1.4em}

\newcommand{\FESTIM}{\textsc{festim}}
\newcommand{\TMAP}{\textsc{tmap}}

\newcommand{\HYPERION}{\textsc{hyperion}}
\newcommand{\Da}{\mathrm{Da}}

\newcommand{\cm}[1]{c^{\mathrm{m}}_{#1}}   
\newcommand{\cs}[1]{c^{\mathrm{s}}_{#1}}   
\newcommand{\Bra}{\mathcal{B}}             

\usepackage{siunitx}
\AtBeginDocument{\let\cm\cmconc}
\journal{Fusion Engineering and Design}

\begin{document}

\begin{frontmatter}

\title{When chemical potential continuity fails:\\
       kinetic interface models for hydrogen isotope transport}

\author[mit]{Remi Delaporte-Mathurin}
\author[mit]{James Dark}

\address[mit]{Massachusetts Institute of Technology, Cambridge, MA, USA}

\begin{abstract}
Macroscopic hydrogen transport codes model material interfaces with local thermodynamic equilibrium (LTE), imposing continuity of chemical potential as a per-species constraint. 
Three assumptions hide in that condition: fast interfacial equilibration, a single exchange pathway between the two sides, and a carrier species known in advance on each side.
The literature scrutinises the first, but the other two are the more consequential failures, and neither survives at a metal/molten-salt interface. 
We replace the constraint with reversible reaction channels at the interface that obey mass action, with detailed balance fixing each ratio of rate constants from the thermodynamic data that already parameterise LTE, and we implement the framework in \FESTIM{}.
LTE is recovered as the fast-kinetics limit of a \emph{single} channel, in both its Sieverts/Sieverts and its Sieverts/Henry form, so the framework generalises LTE and does not compete with it.
A Damk\"ohler number delimits validity within a channel, a branching ratio between channels.
In a representative nickel/FLiBe system, hydrogen partitions kinetically between molecular and fluoride carriers.
The apparent interfacial law then drifts between Sieverts and Henry with loading and salt redox state, and an LTE condition underestimates the steady permeating flux.
The measured pressure exponent is set by the branching ratio, not by any fixed property of the salt; a redox sweep at fixed temperature should continuously shift it between 0.5 and 1.
With two isotopes, two metal-side species feed five salt-side carriers, and a per-species LTE condition is ill-posed.
\end{abstract}


\end{frontmatter}

\section{Introduction}
\label{sec:intro}

Hydrogen isotope transport governs fuel retention, permeation losses, and tritium accountancy in fusion systems~\cite{meschini_impact_2025, meschini_modeling_2023}.
In a liquid-breeder blanket, tritium must cross at least one metal-liquid interface before reaching an extraction system.
Whatever condition is imposed at the interface, therefore, determines how much tritium remains in the liquid, how much escapes through the walls, and how quickly the rest can be recovered.

Macroscopic transport codes such as \FESTIM{}~\cite{dark_festim_2026}, \TMAP{}8~\cite{simon_moose-based_2025} and \textsc{tessim-x}~\cite{schmid_implications_2024} model such interfaces with local thermodynamic equilibrium (LTE), that is, with continuity of the chemical potential of the transported species.
In practice, this becomes an algebraic constraint of Dirichlet type, imposed per species, relating the two interfacial concentrations~\cite{hodille_tritium_2026, shimada_toward_2024, ferrero_preliminary_2022, dark_influence_2021, arredondo_preliminary_2021, delaporte-mathurin_influence_2021, ogorodnikova_model_2002}: it reads $c_A/K_{S,A} = c_B/K_{S,B}$ where hydrogen dissolves dissociatively on both sides, and $c_B = K_{H,B}\,(c_A/K_{S,A})^2$ where it dissolves dissociatively on one side and molecularly on the other.

Three assumptions are hidden in this constraint, and they are rarely stated together.
(A1) is that interfacial equilibration is fast compared with bulk transport.
(A2) is that a \emph{single} exchange pathway connects the two sides.
(A3) is that the carrier species on each side is known a priori and is fixed, so that one solubility law applies throughout.

Of the three, (A1) has been scrutinised.
Molecular dynamics at the Be/BeO interface shows that LTE is not reached even at \qty{1500}{\kelvin}~\cite{hodille_molecular_2022}, and a combined density-functional and rate-equation treatment of W/Cu shows that the steady state reached at such an interface is not the thermodynamic equilibrium state whenever a net flux is carried at that interface~\cite{silva-solis_hydrogen_2026}.

Assumptions (A2) and (A3) have received almost no attention, and we argue that they are the more consequential failures.
The system that motivates the argument is a metal in contact with a fluoride melt, where neither assumption survives the chemistry.

Hydrogen leaving a metal lattice into a molten fluoride has more than one chemical fate.
It may recombine to $\mathrm{H_2}$ and dissolve physically, obeying Henry's law, or the melt may oxidise it to HF and dissolve chemically.
The branching between the two is based on the kinetics of the different pathways, and it moves with the redox state of the salt~\cite{lam_impact_2021,carotti_electrochemical_2021}.
Assumption (A2) asserts that one of these pathways may be ignored, and the chemistry does not say which.

The ambiguity is not hypothetical.
Measuring tritium permeation through a nickel membrane into FLiBe, Calderoni et al.~\cite{calderoni_measurement_2008} concluded that tritium absorbed atomically in the nickel does \emph{not} simply recombine at the Ni/FLiBe interface, and could not distinguish transport as T bound to $\mathrm{BeF_4^{2-}}$ from transport as HT, or from a mixture of the two.
The speciation at that interface is still an open question~\cite{delaporte-mathurin_baby_2025, delaporte-mathurin_advancing_2024}, and reported FLiBe transport properties for $\mathrm{H_2}$ and HF, and for their isotopologues, still span orders of magnitude~\cite{ferry_libra_2022}.

The ambiguity has already contaminated the literature data, because \emph{a solubility constant has no law-independent units}.
Sieverts-type dissolution gives \unit{\mole\per\cubic\metre\pascal\tothe{-1/2}} and Henry-type gives \unit{\mole\per\cubic\metre\per\pascal}, and no fixed factor converts one into the other.
Reporting a number, therefore, \textit{presupposes} the speciation that is itself the quantity in question (assumption (A3)).
Section~\ref{sec:units} expands on this point, as a motivating symptom and not as a criticism of any measurement.

With more than one isotope, (A3) fails structurally.
Two mobile atomic species in the metal, H and T, feed five carriers in the salt, $\mathrm{H_2}$, HT, $\mathrm{T_2}$, HF and TF. 
The salt concentration $c_\mathrm{HT}$ depends on the \emph{product} of the two metal-side concentrations, and no per-species condition on the continuity of a chemical potential can express that. 
LTE here is not inaccurate; it is ill-posed.

Every ingredient of a kinetic interface condition can be found in the literature.
Finite-rate surface kinetics is used for gas/solid interfaces~\cite{ali-khan_rate_1978,pick_model_1985}, and detailed balance is the usual constraint tying forward and reverse constants in a surface reaction network~\cite{mhadeshwar_thermodynamic_2003,gorban_extended_2011}. 
The geochemistry community abandoned local equilibrium for transport-coupled interfacial chemistry three decades ago, deciding when it may nonetheless be used with a Damk\"ohler number that compares the interfacial reaction rate with the rate at which transport supplies it~\cite{knapp_spatial_1989,steefel_coupled_1994}.
However, no macroscopic hydrogen transport code models a condensed/condensed (eg. metal/metal, metal/liquid) interface with competing reaction channels and multiple isotopologues.

We close this gap by formulating a general kinetic interface framework in which the algebraic (Dirichlet-type) constraint is replaced by reversible reaction channels obeying mass action law, with detailed balance fixing the ratio of each pair of rate constants from the same thermodynamic data that parameterise LTE, and we implement it in the open-source finite element code \FESTIM{}.
We recover LTE analytically as the fast-kinetics limit of a single channel, in both its Sieverts/Sieverts and Sieverts/Henry forms, so that what we propose is a strict generalisation of LTE rather than a competing model.
Two dimensionless groups delimit its validity: a Damk\"ohler number within a channel and a branching ratio between channels.
Finally, we apply the framework to the \HYPERION{} Ni/FLiBe/gas experiment~\cite{saraswat_permeation_2026}, in a single-isotope configuration.

Interfaces are treated in this paper as reactive surfaces without their own stored inventory.
Intermediate trapping sites localised on the interface (as done in other studies~\cite{silva-solis_hydrogen_2026, hodille_molecular_2022}) and the codimension-1 formulation needed to carry an interfacial ODE/PDE alongside the bulk equations are the subject of a future paper.

All simulation code, input files and the scripts that produce every figure in this paper are openly available~\cite{remi_delaporte_mathurin_2026_22112697}.

\section{Interface conditions for hydrogen transport}
\label{sec:models}

In this section, we consider two subdomains $\Omega_A$ and $\Omega_B$ that share an interface, $\Gamma$.
In each subdomain, the transport of a dissolved species obeys the usual diffusion equation $\partial_t c = \nabla\cdot(D\nabla c) + S$ solved by \FESTIM{}. 
Trapping is omitted, since it plays no part in the arguments that follow.
The models below differ only in how the two bulk problems are coupled across $\Gamma$.

\subsection{Local thermodynamic equilibrium and its assumptions}
\label{sec:lte}

Local thermodynamic equilibrium (LTE) is the statement that the chemical potential of dissolved hydrogen is continuous across $\Gamma$.
Concentration is not, since the relation between the two depends on the dissolution law obeyed on each side, so the algebraic condition that follows takes a different form for each pair of laws. 
For two materials in which hydrogen dissolves dissociatively, Sieverts' law on both sides, it reads:
\begin{equation}
    \frac{c_A\big|_\Gamma}{K_{S,A}(T)}
    = \frac{c_B\big|_\Gamma}{K_{S,B}(T)}
    \quad \text{(Sieverts / Sieverts)},
    \label{eq:lte_ss}
\end{equation}
whereas for a solid in contact with a liquid in which hydrogen dissolves molecularly (Henry's law), it becomes:
\begin{equation}
    c_B\big|_\Gamma
    = K_{H,B}(T)\left(\frac{c_A\big|_\Gamma}{K_{S,A}(T)}\right)^{2}
    \quad \text{(Sieverts / Henry)} .
    \label{eq:lte_sh}
\end{equation}

Both conditions are \emph{algebraic}: they fix a ratio, or a power law, between the two interfacial concentrations, whatever the flux the interface is carrying.
Three consequences follow, and they are properties of the constraint, not of the system to which it is applied. 
The interface has no timescale of its own: it responds instantaneously to any change in the adjacent bulk fields.
It offers no resistance, so $\Gamma$ can never be the rate-limiting step of a permeation problem.
And, the exponent distinguishing Eq.~(\ref{eq:lte_ss}) from Eq.~(\ref{eq:lte_sh}), and hence the identity of the carrier species on the $B$ side, must be chosen before the simulation is run.

LTE also contradicts itself at a permeating interface, as noted by Silva-Sol\'is et al.~\cite{silva-solis_hydrogen_2026}.
Thermodynamic equilibrium requires every microscopic forward rate to be balanced by its reverse, which implies \emph{zero} net flux across $\Gamma$. 
A permeation experiment, however, carries a nonzero net flux by construction.
Imposing LTE at a permeating interface is therefore always an approximation.
The only question is how poor the approximation is, and Sec.~\ref{sec:damkohler} makes that question quantitative.

Two notions are routinely conflated and are kept distinct throughout.
A \emph{steady state} satisfies $\partial_t c = 0$ with $J_\mathrm{in} = J_\mathrm{out} \neq 0$, whereas \emph{equilibrium} satisfies $\partial_t c = 0$ with $J_\mathrm{in} = J_\mathrm{out} = 0$.
Every equilibrium is a steady state, but a steady state carrying a flux is not an equilibrium; that is what LTE discards.

\subsection{A general kinetic interface framework}
\label{sec:general}

We replace the algebraic constraint by a set of reversible reaction channels living on $\Gamma$. Let $X_i$ denote the mobile atomic species on side $A$ ($i \in \{\mathrm{H,D,T}\}$) and $Y_\alpha$ the carriers on side $B$ ($\alpha \in \{\mathrm{H_2, HT, T_2, HF, TF},\dots\}$).
A channel $r$ is written:
\begin{equation}
    \sum_i \nu_{ir} X_i + \sum_\beta \lambda_{\beta r} Z_\beta
    \;\rightleftharpoons\; \sum_\alpha \mu_{\alpha r} Y_\alpha,
    \label{eq:channel}
\end{equation}
where the $Z_\beta$ are non-hydrogenic constituents entering the stoichiometry, the fluoride ion and the oxidising or reducing half of the salt redox buffer, in the application of Sec.~\ref{sec:model3}. 
We represent them by activities $a_\beta$, not by transported fields. 
Each channel proceeds at a net rate given by mass action,
\begin{equation}
    \begin{split}
        w_r = {}& k_r^{+}(T)\prod_i \left(\cm{i}\big|_\Gamma\right)^{\nu_{ir}}
                  \prod_\beta a_\beta^{\lambda_{\beta r}} \\
              &- k_r^{-}(T)\prod_\alpha \left(\cs{\alpha}\big|_\Gamma\right)^{\mu_{\alpha r}} ,
    \end{split}
    \label{eq:mass_action}
\end{equation}
expressed per unit interfacial area, so that $w_r$ has units of \unit{\mole\per\square\metre\per\second} and the units of $k_r^{\pm}$ depend on the order of the channel: an exchange velocity in \unit{\metre\per\second} for a first-order channel, \unit{\metre\tothe{4}\per\mole\per\second} for a second-order one.
The interface conditions then state that the net atomic flux of isotope $i$ leaving $A$, and the production of carrier $\alpha$ into $B$, are sums over all channels.
Throughout, $\mathbf{n}$ denotes the outward normal of the subdomain whose equation is being written:
\begin{align}
    -D^{\mathrm{m}}_i \nabla \cm{i}\cdot \mathbf{n}\big|_\Gamma
        &= \sum_r \nu_{ir}\, w_r
    \label{eq:bc_metal}\\
     D^{\mathrm{s}}_\alpha \nabla \cs{\alpha}\cdot \mathbf{n}\big|_\Gamma
        &= \sum_r \mu_{\alpha r}\, w_r.
    \label{eq:bc_salt}
\end{align}

Conservation of atoms across $\Gamma$ is not an additional requirement but a condition on the stoichiometric coefficients.
Writing $n_{i\alpha}$ for the number of atoms of isotope $i$ carried by species $\alpha$, every channel must satisfy
\begin{equation}
    \nu_{ir} = \sum_\alpha \mu_{\alpha r}\, n_{i\alpha}
    \qquad \text{for each isotope } i,
    \label{eq:atom_balance}
\end{equation}
so that the atomic flux removed from $A$ by Eq.~(\ref{eq:bc_metal}) is exactly the atomic flux delivered to $B$ by Eq.~(\ref{eq:bc_salt}). 
For the recombination channel of Sec.~\ref{sec:model2}, for instance,
$\nu_\mathrm{H} = 2$ and $\mu_{\mathrm{H_2}} = 1$ with $n_{\mathrm{H,H_2}} = 2$.

The forward and reverse rate constants of each channel are not independent.
Detailed balance requires
\begin{equation}
    \frac{k_r^{+}(T)}{k_r^{-}(T)} = K_r(T),
    \label{eq:detailed_balance_general}
\end{equation}
where $K_r$ is the equilibrium constant of that channel.
For every channel in this paper, $K_r$ follows from the solubilities of the participating species.
Thermodynamics therefore fixes the \emph{ratio} of the rate constants.
Their \emph{magnitude}, how fast the interface equilibrates, must come from atomistic modelling or from experiments.

The framework is therefore a strict generalisation of LTE: the same thermodynamic data that parameterise Eqs.~(\ref{eq:lte_ss}) and~(\ref{eq:lte_sh}) parameterise Eq.~(\ref{eq:mass_action}), and the one extra input is the magnitude, which LTE fixes at infinity without saying so.
Eq~(\ref{eq:detailed_balance_general}) constrains whoever writes the model and cannot be enforced by the discretisation~\cite{mhadeshwar_thermodynamic_2003, mhadeshwar_thermodynamically_2005, gorban_extended_2011}; Sec.~\ref{sec:params} returns to how the ratios are obtained in practice.

\begin{figure*}[ht]
    \centering
    \includegraphics[width=0.8\linewidth]{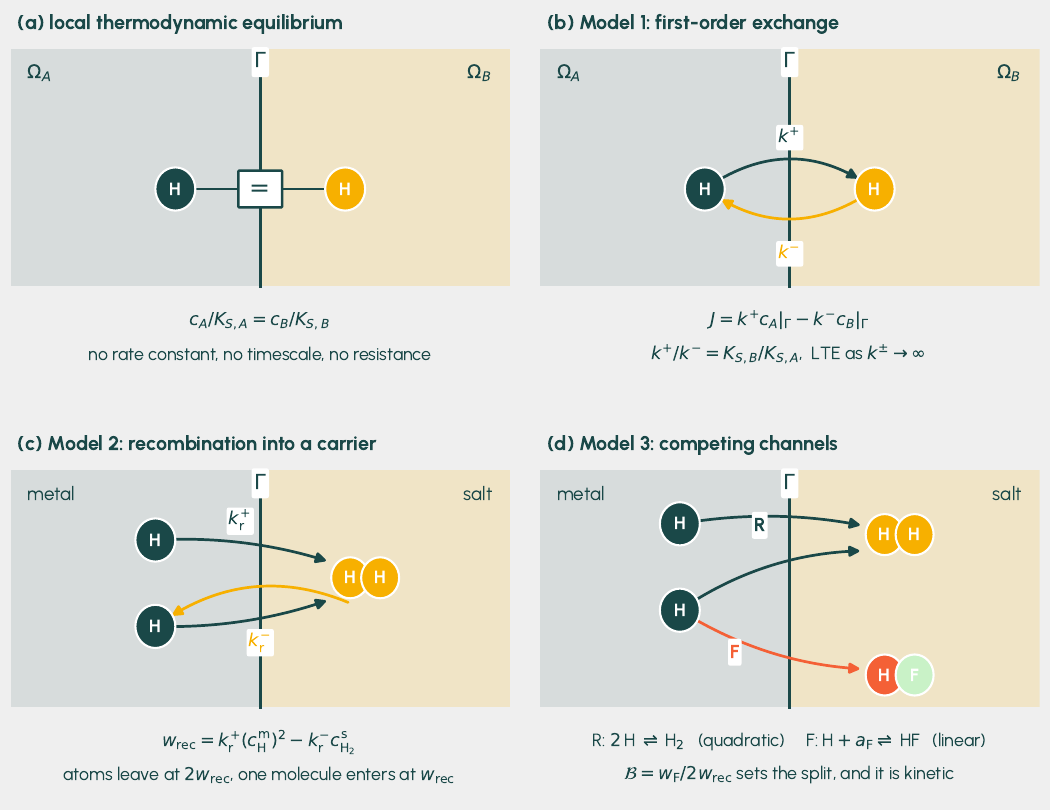}
    \caption{The four interface models on the same two-subdomain geometry, differing only in the chemistry allowed on $\Gamma$. (a) LTE: nothing crosses, the two interfacial concentrations being tied by the algebraic constraint of Eq.~(\ref{eq:lte_ss}), which gives the interface no rate constant, no timescale and no resistance. (b) Model 1, Eq.~(\ref{eq:model1_flux}): an atom crosses without changing chemical identity, first order in either direction. (c) Model 2, Eq.~(\ref{eq:model2_rate}): two metal-side atoms recombine into one molecular carrier, so the atomic flux leaving the metal is twice the channel rate, Eq.~(\ref{eq:atom_balance}). (d) Model 3: the same atom has two exits of different order, recombination (quadratic) and fluorination (linear), and the branching ratio $\Bra$ of Eq.~(\ref{eq:branching}) sets how the atomic flux divides between them.} 
    \label{fig:interface_models}
\end{figure*}

The framework is agnostic to the nature of the two sides: the gas/solid
dissociation-recombination boundary condition already available in \FESTIM{}~\cite{ali-khan_rate_1978, pick_model_1985, hodille_kinetic_2020, kulagin_kinetic_2024} is the special case in which $A$ is a gas phase.
The models below differ from one another only in their stoichiometry and in the number of channels they carry, and Fig.~\ref{fig:interface_models} collects them on the same geometry, together with the LTE condition they generalise.

\subsection{Model 1: first-order exchange}
\label{sec:model1}

The simplest case is a single channel $X_A \rightleftharpoons X_B$ in which an atom crosses the interface without changing chemical identity, as at a metal/metal interface (see Fig.~\ref{fig:interface_models}b).
Eq~(\ref{eq:mass_action}) then reduces to a Robin-type flux condition,
\begin{equation}
    J = k^{+} c_A\big|_\Gamma - k^{-} c_B\big|_\Gamma ,
    \label{eq:model1_flux}
\end{equation}
with both rate constants an exchange velocity in \unit{\metre\per\second}, and with mass conservation across $\Gamma$ giving $-D_A \,\partial_n c_A = D_B\, \partial_n c_B = J$.
Setting $J = 0$ and requiring the resulting concentration ratio to match Eq.~(\ref{eq:lte_ss}) fixes
\begin{equation}
    \frac{k^{+}}{k^{-}} = \frac{K_{S,B}(T)}{K_{S,A}(T)},
    \label{eq:detailed_balance_1}
\end{equation}
which is Eq.~(\ref{eq:detailed_balance_general}) for this channel. 
Writing $K \equiv k^{+}/k^{-}$ and factorising Eq.~(\ref{eq:model1_flux}) as $J = k^{+}\left(c_A\big|_\Gamma - c_B\big|_\Gamma/K\right)$, a finite flux at $k^{+} \rightarrow \infty$ forces the bracket to vanish, which is Eq.~(\ref{eq:lte_ss}) exactly.
The derivation is given in \ref{app:limits}. 
The correction at finite rate constants is $\mathcal{O}(J/k^{+}c)$ and is quantified in Sec.~\ref{sec:damkohler}.

An interfacial rate constant in series with a bulk diffusive resistance is familiar in the solid state; it is for instance used to describe oxidation laws~\cite{deal_general_1965} and the Kapitza resistance for heat (also known as \textit{interfacial thermal resistance)}~\cite{swartz_thermal_1989}. 
Interstitial diffusion is itself a first-order exchange between adjacent planes of sites, so Fick's law is the special case $k^{+} = D/\lambda$ with $\lambda$ the jump distance.

\subsection{Model 2: recombination into a molecular carrier}
\label{sec:model2}

The second case retains a single channel but changes the chemical identity of the transported species, $2\,X_A \rightleftharpoons Y_{A_2}$: atomic hydrogen dissolved in the metal recombines into a molecule that dissolves physically in the liquid (see Fig.~\ref{fig:interface_models}c).
It is the metal/molten-salt (denoted as m and s, respectively) analogue of surface recombination, and its net rate is
\begin{equation}
    w_{\mathrm{rec}} = k_{\mathrm{r}}^{+}\left(\cm{\mathrm{H}}\big|_\Gamma\right)^{2}
                     - k_{\mathrm{r}}^{-}\,\cs{\mathrm{H_2}}\big|_\Gamma ,
    \label{eq:model2_rate}
\end{equation}
with interface conditions $-D^{\mathrm{m}}\partial_n \cm{\mathrm{H}} = 2 w_{\mathrm{rec}}$ and $D^{\mathrm{s}}\partial_n \cs{\mathrm{H_2}} = w_{\mathrm{rec}}$, the factor two being the stoichiometry of Eq.~(\ref{eq:atom_balance}). 
Imposing $w_{\mathrm{rec}} = 0$ together with Sieverts' law in the metal ($c^{\mathrm{m}} = K_S\sqrt{P}$) and Henry's law in the liquid ($c^{\mathrm{s}} = K_H P$) gives
\begin{equation}
    \frac{k_{\mathrm{r}}^{+}}{k_{\mathrm{r}}^{-}}
    = \frac{K_{H}(T)}{K_{S}(T)^{2}} .
    \label{eq:detailed_balance_2}
\end{equation}
Substituting Eq.~(\ref{eq:detailed_balance_2}) back into $w_{\mathrm{rec}} = 0$ returns the Sieverts/Henry condition, Eq.~(\ref{eq:lte_sh}), exactly.
So, the mixed-law interface currently available in \FESTIM{}~v2 is the fast-kinetics limit of Eq.~(\ref{eq:model2_rate}).
The two interface laws of Sec.~\ref{sec:lte} are therefore not separate physical models to be selected between, but the fast limits of two different channels of the same framework.
Which one applies is a question about chemistry, not about which option to set in an input file.

\subsection{Model 3: competing channels}
\label{sec:model3}

Nothing in Sec.~\ref{sec:general} restricts the interface to a single channel, and at a metal/fluoride-melt interface there is no reason to expect one. 
Hydrogen leaving the metal may recombine and dissolve as a molecule, or the melt may oxidise it to HF, which is far more soluble, corrosive, and redox-dependent.
We therefore let two channels operate on the same interface (see Fig.~\ref{fig:interface_models}d),
\begin{align*}
    \text{R:}&\quad 2\,\mathrm{H(m)} \rightleftharpoons \mathrm{H_2(s)} \\
    \text{F:}&\quad \mathrm{H(m)} + \mathrm{F^-(s)} + h^{+}
                    \rightleftharpoons \mathrm{HF(s)} ,
\end{align*}
where $h^{+}$ denotes the oxidising half of the salt redox couple.
In practice, the fluorine potential is set by a buffer, either beryllium metal or an imposed $\mathrm{HF/H_2}$ ratio in the cover gas, so we write channel F with a single effective activity, $a_\mathrm{F}$, lumping the fluoride activity and the redox potential,
\begin{equation}
    w_{\mathrm{F}} = k_{\mathrm{f}}^{+}\, a_{\mathrm{F}}\,
                     \cm{\mathrm{H}}\big|_\Gamma
                   - k_{\mathrm{f}}^{-}\,\cs{\mathrm{HF}}\big|_\Gamma ,
    \label{eq:model3_rate}
\end{equation}
while $w_\mathrm{rec}$ retains the form of Eq.~(\ref{eq:model2_rate}). 
The atomic flux leaving the metal is now shared between the two,
\begin{equation}
    -D^{\mathrm{m}}\nabla \cm{\mathrm{H}}\cdot\mathbf{n}\big|_\Gamma
    = 2\,w_{\mathrm{rec}} + w_{\mathrm{F}} ,
    \label{eq:model3_total_flux}
\end{equation}
and each channel feeds its own carrier in the salt through Eq.~(\ref{eq:bc_salt}).

The two channels are of different order in the interfacial loading: R is quadratic in $\cm{\mathrm{H}}$ and F is linear.
Their competition is measured by the branching ratio
\begin{equation}
    \Bra
    \equiv \frac{w_{\mathrm{F}}}{2\,w_{\mathrm{rec}}}
    \;\xrightarrow[\text{far from equilibrium}]{}\;
    \frac{k_{\mathrm{f}}^{+} a_{\mathrm{F}}}
         {2\,k_{\mathrm{r}}^{+}\,\cm{\mathrm{H}}\big|_\Gamma} ,
    \label{eq:branching}
\end{equation}
whose consequences form the core of this paper's argument.

First, $\Bra$ depends \emph{inversely} on the interfacial loading.
At low loading, the fluorination channel dominates and the atomic flux crossing the interface scales linearly with $\cm{\mathrm{H}}\big|_\Gamma$; at high loading, recombination takes over, and the scaling becomes quadratic. 
Defining the apparent interfacial exponent on that flux, the right-hand side of Eq.~(\ref{eq:bc_metal}) for $i = \mathrm{H}$,
\begin{equation}
    n \equiv \frac{\partial \ln J}{\partial \ln \cm{\mathrm{H}}\big|_\Gamma} ,
    \quad J \equiv \sum_r \nu_{\mathrm{H}r}\, w_r ,
    \quad 1 \le n \le 2 ,
    \label{eq:apparent_exponent}
\end{equation}
LTE requires $n$ to be a constant, fixed in advance: $n = 2$ for the Sieverts/Henry condition of Eq.~(\ref{eq:lte_sh}), $n = 1$ for a Henry/Henry-like one.
Here $n$ is instead a solution-dependent quantity that drifts during a transient as the interfacial loading builds up.
No fixed algebraic interface law can reproduce this behaviour.

The same slope can be read off the salt-side inventory instead of the flux, as a fitted solubility does, and the two are not the same number unless the carriers are transported alike (\ref{app:analytical}, Sec.~\ref{sec:exponent_verification}).
We take the flux as the definition throughout, since it is what a permeation experiment measures directly.

Second, $\Bra$ depends on the salt redox state through $a_\mathrm{F}$, which is itself dynamic: HF production, corrosion of the container and depletion of the buffer all move it. 
The behaviour of the interface is therefore coupled to the salt chemistry and is not a material property of the metal/salt pair that could be tabulated once and reused.

Third, fast kinetics does not rescue LTE here.
Even when both channels are individually equilibrated, with a large Damk\"ohler number for each in the sense of Sec.~\ref{sec:damkohler}, the \emph{partition} of the flux between them is set by the ratio of the forward rate constants and not by thermodynamics.
Making each channel faster makes each channel individually closer to its own equilibrium while leaving the split between them kinetic.

It is possible, in principle, to treat the interface without any kinetic information by imposing full local chemical equilibrium among all interfacial species, a Gibbs minimisation at $\Gamma$ with the prescribed fluorine potential.
That description is, however, (i) not what any macroscopic hydrogen transport code implements; (ii) valid only if every interconversion is fast compared with transport, which is exactly what the available evidence puts in doubt~\cite{hodille_molecular_2022, calderoni_measurement_2008}; and (iii) still silent on how a given net flux divides between carriers, since equilibrium fixes concentrations and not fluxes.
The last point is the decisive one for the quantity measured in a permeation experiment.

\subsection{Multiple isotopes: isotopologue channels}
\label{sec:isotopes}

The framework extends to $N_\mathrm{iso}$ mobile atomic species in the metal without modification, but the structure that emerges is qualitatively new.
Recombination now populates every isotopologue and fluorination every fluoride. 
The gas/solid case has already met this structure: co-permeation of hydrogen isotopes through a metal forms the mixed molecule at a rate set by the product of the two surface concentrations~\cite{zhu_modeling_2024}.
Dropping $\big|_\Gamma$ for readability, for $\{\mathrm{H,T}\}$ the recombination channels are\footnote{The rate laws carry no combinatorial prefactor, so the statistical degeneracy of the mixed pair sits in the value of the constant: $k_{\mathrm{HT}}^{+} = 2\,k_{\mathrm{HH}}^{+} = 2\,k_{\mathrm{TT}}^{+}$ in the mass-independent limit, with equal reverse constants, which returns the classical $K_\mathrm{exch} = (\cs{\mathrm{HT}})^{2} / (\cs{\mathrm{H_2}}\,\cs{\mathrm{T_2}}) = 4$, random pairing of equal H and T populations.}
\begin{align}
    w_{\mathrm{HH}} &= k_{\mathrm{HH}}^{+}\left(\cm{\mathrm{H}}\right)^{2}
                     - k_{\mathrm{HH}}^{-}\,\cs{\mathrm{H_2}}
    \label{eq:iso_HH}\\
    w_{\mathrm{HT}} &= k_{\mathrm{HT}}^{+}\,\cm{\mathrm{H}}\,\cm{\mathrm{T}}
                     - k_{\mathrm{HT}}^{-}\,\cs{\mathrm{HT}}
    \label{eq:iso_HT}\\
    w_{\mathrm{TT}} &= k_{\mathrm{TT}}^{+}\left(\cm{\mathrm{T}}\right)^{2}
                     - k_{\mathrm{TT}}^{-}\,\cs{\mathrm{T_2}},
    \label{eq:iso_TT}
\end{align}
and the fluorination channels are
\begin{align}
    w_{\mathrm{HF}} &= k_{\mathrm{f,H}}^{+} a_{\mathrm{F}} \cm{\mathrm{H}}
                     - k_{\mathrm{f,H}}^{-}\,\cs{\mathrm{HF}}
    \label{eq:iso_HF}\\
    w_{\mathrm{TF}} &= k_{\mathrm{f,T}}^{+} a_{\mathrm{F}} \cm{\mathrm{T}}
                     - k_{\mathrm{f,T}}^{-}\,\cs{\mathrm{TF}} .
    \label{eq:iso_TF}
\end{align}

The metal-side conditions follow from Eq.~(\ref{eq:bc_metal}) with the stoichiometry of Eq.~(\ref{eq:atom_balance}),
\begin{align}
    -D^{\mathrm{m}}_\mathrm{H}\nabla \cm{\mathrm{H}}\cdot\mathbf{n}\big|_\Gamma
        &= 2\,w_{\mathrm{HH}} + w_{\mathrm{HT}} + w_{\mathrm{HF}}
    \label{eq:iso_flux_H}\\
    -D^{\mathrm{m}}_\mathrm{T}\nabla \cm{\mathrm{T}}\cdot\mathbf{n}\big|_\Gamma
        &= 2\,w_{\mathrm{TT}} + w_{\mathrm{HT}} + w_{\mathrm{TF}} ,
    \label{eq:iso_flux_T}
\end{align}
each salt-side carrier being fed by the single channel that produces it.

All five channels are written with the plain mass-action form of Eq.~(\ref{eq:mass_action}), and stoichiometry enters only through the coefficients $\nu_{ir}$ of Eq.~(\ref{eq:bc_metal}).

Counting the carriers makes the multi-isotope case structurally different.
Two mobile species on the metal side feed five carriers on the salt side, whereas LTE supplies one scalar constraint per transported species pair. 

With two metal-side and five salt-side interfacial unknowns, the per-species Dirichlet condition is underdetermined.
The more serious objection is that it is also wrong in form: HT is made from one H and one T, so by Eq.~(\ref{eq:iso_HT}) its interfacial concentration depends on both metal-side concentrations at once. 
A condition written one species at a time has nothing to equate it to.

Isotope exchange, $\mathrm{H_2} + \mathrm{T_2} \rightleftharpoons 2\,\mathrm{HT}$, is slow in the homogeneous phase but is catalysed by metal surfaces. 
In the present framework, it requires no additional reaction: it emerges from Eqs.~(\ref{eq:iso_HH})--(\ref{eq:iso_TT}), since a $\mathrm{T_2}$ molecule may dissociate at $\Gamma$ and deposit T into the lattice, and lattice T may subsequently leave paired with an H.
Interfacial scrambling is thus predicted by the model, not imposed on it.

Two observable consequences follow, neither of which can be expressed by LTE.
The first is \emph{isotope swamping}: the tritium flux depends on the protium inventory because raising $\cm{\mathrm{H}}$ shifts tritium from the $\mathrm{T_2}$ channel to the statistically favoured HT channel, thereby changing the total tritium throughput.
A per-species condition predicts no such coupling, since its tritium condition does not contain the protium concentration.
The second is \emph{interfacial fractionation}: because the recombination and fluorination channels carry different isotope dependences ($k_\mathrm{HH}$ against $k_\mathrm{HT}$ against $k_\mathrm{TT}$; $k_\mathrm{f,H}$ against $k_\mathrm{f,T}$), the isotopic composition of the permeating flux differs from that of the metal, and differs between the molecular and the fluoride carrier. 
The model outputs $\alpha = (\mathrm{T/H})_\mathrm{flux} / (\mathrm{T/H})_\mathrm{metal}$.

Detailed balance applies channel by channel as in Eq.~(\ref{eq:detailed_balance_general}), the ratios $k^{+}/k^{-}$ being constrained by the isotopologue equilibrium constants. 
These are close to, but not exactly, the classical statistical values.

\subsection{Dimensionless criteria for LTE validity}
\label{sec:damkohler}

The framework of Sec.~\ref{sec:general} contains LTE as a limit, so it can also say when that limit is legitimate. 
Three questions decide this, and we consider them in increasing order of importance: whether a channel equilibrates fast enough, whether a single channel dominates, and whether each isotope reaches the salt on its own.

\paragraph{Within a channel} 
The relevant comparison is between interfacial equilibration and bulk transport, and a dimensionless group based on that ratio is the standard test of the local equilibrium assumption~\cite{knapp_spatial_1989}.
In gas-driven permeation through metals, this is known as the surface-limited regime~\cite{pisarev_gas-driven_2003, denisov_surface-limited_2018}.
Because channels may be of different order, the comparison uses the exchange velocity obtained by linearising the channel about the interfacial concentration, $k_r^\mathrm{eff} \equiv \partial w_r / \partial \cm{}\big|_\Gamma$, which reduces to $k^{+}$ for Model 1 and to $2 k_\mathrm{r}^{+} \cm{\mathrm{H}}|_\Gamma$ for Model 2.
The Damk\"ohler number of channel $r$ is then
\begin{equation}
    \Da_r = \frac{k_r^\mathrm{eff}\,L}{D} .
    \label{eq:damkohler}
\end{equation}
When $\Da_r \gg 1$, the channel is locally equilibrated, and its algebraic LTE form is recovered, as shown in Secs.~\ref{sec:model1} and~\ref{sec:model2}. 
For asymmetric systems, one Damk\"ohler number should be evaluated per side, using $D_A, L_A$ and $D_B, L_B$ respectively, and the smaller of the two governs.

For a channel of order one, as in Model 1, $k^\mathrm{eff}$ is a constant and $\Da$ is an input to the problem.
For Model 2, neither is true: the metal side sees an exchange velocity $2k_\mathrm{r}^{+}\cm{\mathrm{H}}|_\Gamma$, which falls as the interface is depleted.
This means that, for instance, in a permeation transient, the same interface may be kinetically limited initially (i.e., when the interfacial concentration is low) and then effectively equilibrates at high interfacial concentration.
A sweep in Da therefore needs a reference concentration.
We use the upstream Sieverts value $c^{\star} = K_S\sqrt{P_\mathrm{up}}$, giving the control parameter $\Da^{\star} = 2k_\mathrm{r}^{+}c^{\star}L/D$, and report the value attained at the interface as a diagnostic.
That reference is known from the boundary condition before anything is solved, and every dimensionless group quoted in this paper is built at it, the branching ratio included, so both axes of the map below refer to one concentration.
Which concentration a Damk\"ohler number is built with is part of its definition and should be quoted with it.

The caveat carried over from Ref.~\cite{silva-solis_hydrogen_2026} applies throughout: $\Da \gg 1$ is \emph{necessary but not sufficient}, since under a net flux the steady state departs from equilibrium by a term proportional to $J/k^{+}$, which LTE sets to zero by construction.


\paragraph{Between channels}
The relevant quantity is the branching ratio $\Bra$ of Eq.~(\ref{eq:branching}).
As $\Bra \rightarrow 0$ recombination dominates, and the interface obeys the Sieverts/Henry condition of Eq.~(\ref{eq:lte_sh}).
At $\Bra \rightarrow \infty$ fluorination dominates, and the interface law is linear with the metal-side concentration.
Either limit has an LTE form.
In between, both channels carry a substantial share, and no single algebraic law applies at all, no matter how large the Damk\"ohler numbers are.

\begin{figure}[htbp]
    \centering
    \includegraphics[width=1\linewidth]{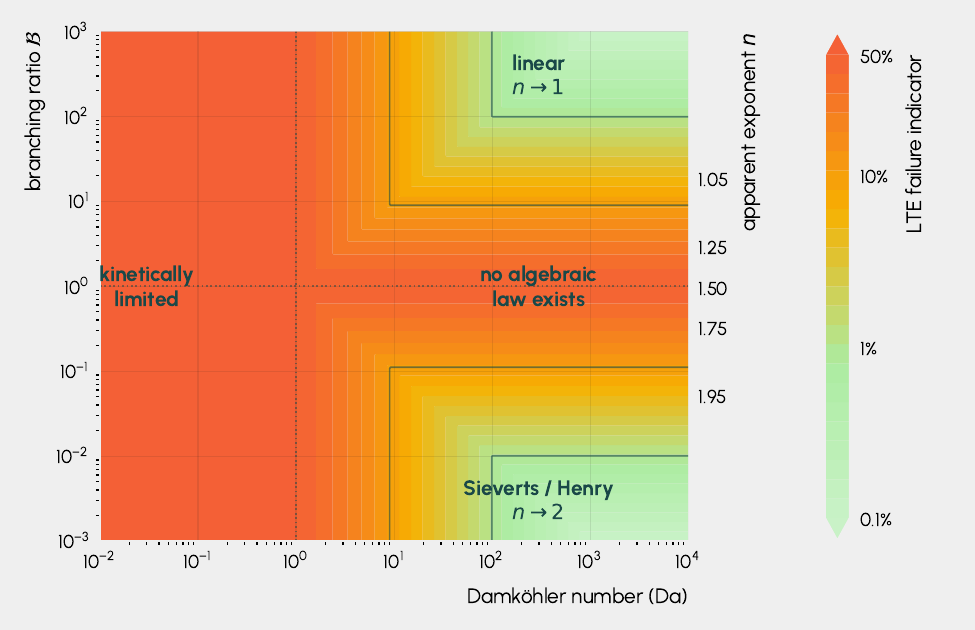}
    \caption{Where a local-equilibrium interface is defensible. 
    The indicator $\mathcal{E}$ of Eq.~(\ref{eq:lte_indicator}) over the $(\Da, \Bra)$ plane,     with contours at \qty{1}{\percent} and \qty{10}{\percent}. 
    The Damk\"ohler number plotted is the two-sided group $\Da^{\star}$ of Eq.~(\ref{eq:defect_1_explicit}), the ratio of the total bulk resistance to the interfacial resistance $1/k^{+}$, since that is the combination the flux error depends on.
    LTE requires both a fast channel and a dominant one: the pale regions at bottom right and top right are the Sieverts/Henry and linear limits, the left-hand strip is kinetically limited whatever the branching, and the band around $\Bra = 1$ admits no algebraic interface law however fast the kinetics.
    The right-hand axis gives the apparent exponent $n$ corresponding to each branching ratio.}
    \label{fig:regime_map}
\end{figure}

$\Da$ and $\Bra$ are independent of one another, and are placed on the two axes of a single map (see Fig.~\ref{fig:regime_map}).
The field plotted is the larger of two quantities,
\begin{equation}
    \mathcal{E} = \max\left(
        \underbrace{\frac{1}{1 + \Da}}_{\text{within a channel}} ,\;
        \underbrace{\frac{\min(1, \Bra)}{1 + \Bra}}_{\text{between channels}}
    \right),
    \label{eq:lte_indicator}
\end{equation}
the first is the relative error made on the flux by treating a channel as equilibrated, exact for Model 1 by Eq.~(\ref{eq:model1_convergence}), and the second is the fraction of the atomic flux carried by the minority channel, the part of the flux that no single-channel model represents at all.
Taking the larger of the two is a convention: $\mathcal{E}$ indicates how badly the best available LTE condition performs.
The map is nevertheless entirely analytical, and the right-hand axis carries the apparent exponent $n = (2+\Bra)/(1+\Bra)$ of Eq.~(\ref{eq:n_of_B}), the quantity a permeation experiment would report.

Read as a decision aid, the map says that an LTE interface needs $\Da^{\star} \gtrsim 100$ \emph{and} a branching ratio at least two orders of magnitude away from unity.
Neither condition alone is sufficient.

\paragraph{Across isotopes} 
LTE further requires that no salt species be fed by two metal-side isotopes at once.
The departure is quantified by the fraction of the tritium flux carried by the mixed isotopologue,
\begin{equation}
    \chi_\mathrm{HT} = \frac{w_\mathrm{HT}}{w_\mathrm{HT} + 2 w_\mathrm{TT}}.
    \label{eq:chi_HT}
\end{equation}
A per-isotope description is recovered as $\chi_\mathrm{HT} \rightarrow 0$ and is forbidden when $\chi_\mathrm{HT} = \mathcal{O}(1)$.
The fluorination channels pair no isotopes, since each feeds its own fluoride, so $\chi_\mathrm{HT}$ is built on the recombination channels alone and is independent of the branching ratio.
In the tritium-lean, protium-rich conditions typical of a breeding blanket, $\chi_\mathrm{HT} \rightarrow 1$.

\section{Implementation in \FESTIM}
\label{sec:implementation}

\FESTIM{}~\cite{dark_festim_2026} is an open-source Python framework for hydrogen isotope transport in materials, built on the finite element library DOLFINx~\cite{baratta_dolfinx_2023}.
Version 2.0 divides a computational domain into subdomains, each discretised on its own submesh, so a field is continuous within a subdomain but free to jump across a shared facet.
The mixed-domain functionality of DOLFINx then combines each subdomain's formulation into a blocked formulation.
The interface conditions of Sec.~\ref{sec:models} enter that form as surface terms, in the same way a flux condition enters on an external boundary, and they prescribe the flux crossing $\Gamma$ rather than the concentrations on either side of it.

\subsection{Weak formulation}
\label{sec:weak}

Multiplying the transport equation of species $i$ on $\Omega_A$ by a test function $v \in H^1(\Omega_A)$ and integrating the diffusive term by parts gives the usual bulk residual.
The interface enters only through the part of the boundary integral carried by $\Gamma$.
Substituting Eqs.~(\ref{eq:bc_metal}) and~(\ref{eq:bc_salt}) leaves the surface contributions
\begin{align}
    F^{\mathrm{m}}_i &\supset
    +\int_\Gamma \Big(\sum_r \nu_{ir}\, w_r\Big)\,
        v^{\mathrm{m}}_i \,\mathrm{d}S
    \label{eq:weak_metal}\\
    F^{\mathrm{s}}_\alpha &\supset
    -\int_\Gamma \Big(\sum_r \mu_{\alpha r}\, w_r\Big)\,
        v^{\mathrm{s}}_\alpha \,\mathrm{d}S ,
    \label{eq:weak_salt}
\end{align}
with $w_r$ given by Eq.~(\ref{eq:mass_action}), evaluated on $\Gamma$ using the values of the solution from either side.
The opposite signs are those of the two outward normals: a channel proceeding forwards removes atoms from $\Omega_A$ and delivers carriers to $\Omega_B$.
The two integrals are assembled with an interior-facet measure, $\mathrm{d}S$, whose integration data map each facet to the cell it belongs to on each side, one restriction per subdomain.

No interfacial degrees of freedom are introduced.
Every channel rate is calculated from concentrations the solver already has, so the interface adds no unknowns of its own and the system is the same size as it would be with an LTE interface on the same mesh\footnote{This is a simplification relative to a codimension-1 coupling, where an interfacial inventory is transported along $\Gamma$ and does carry its own degrees of freedom.}.

The stoichiometric coefficients enter as multiplicities. 
The rate $w_r$ is added once to the residual of each occurrence of a species in the channel, so declaring the recombination channel of Eq.~(\ref{eq:model2_rate}) with two H reactants produces both the square in the rate law and the factor two in Eq.~(\ref{eq:bc_metal}).

The assembled system is blocked, with one residual block per subdomain, and the surface integrals of Eqs.~(\ref{eq:weak_metal}) and~(\ref{eq:weak_salt}) are the only terms coupling one block to another.
Each block is differentiated symbolically with respect to each unknown in the coupled system. 
The isotopologue channels produce a coupling that no per-species has.
Since $\partial w_\mathrm{HT}/\partial \cm{\mathrm{H}} = k^{+}_\mathrm{HT}\cm{\mathrm{T}}$, the residual of H depends on the T concentration, so two species on the \emph{same} side of the interface are coupled through it.
Several channels sharing a single interface are treated as separate objects whose contributions sum, and this is how we assemble Model 3 and the five-channel isotopologue set in Sec.~\ref{sec:isotopes}.

\subsection{\FESTIM{} user API}
\label{sec:user_api}

An interface condition is added to a model as an object carrying the two rate constants and the species involved, and the assembly of Sec.~\ref{sec:weak} is handled internally.
A first-order channel, Model 1, uses:

\begin{python}
import festim as F

XA  = F.Species("XA",  subdomains=[A])
XB = F.Species("XB", subdomains=[B])

# X(A) <=> X(B), first order each way
model_1_eg = F.InterfaceReaction(
    id=1,
    subdomains=[A, B],
    reactants=[XA],
    products=[XB],
    k_plus=k_plus,
    k_minus=k_minus,
)
\end{python}

A channel that changes the carrier, stoichiometry is set by repetition in those lists, so the two H reactants of the recombination channel produce both the square in the rate and the factor two in the atomic flux.
Model~3 is two such channels sharing an interface, their contributions adding on $\Gamma$:

\begin{python}
# metal hosts atomic H; salt hosts the carriers
H  = F.Species("H",  subdomains=[metal])
H2 = F.Species("H2", subdomains=[salt])
HF = F.Species("HF", subdomains=[salt])

# R: 2 H(m) <=> H2(s)
recomb_channel = F.InterfaceReaction(
    id=1,
    subdomains=[metal, salt],
    reactants=[H, H],
    products=[H2],
    k_plus=kr_plus,
    k_minus=kr_minus,
)

# F: H(m) <=> HF(s), a_F folded into k_plus
fluorine_channel = F.InterfaceReaction(
    id=1,
    subdomains=[metal, salt],
    reactants=[H],
    products=[HF],
    k_plus=kf_plus * a_F,
    k_minus=kf_minus,
)

\end{python}

The species restricted to \pyth{[metal]} or \pyth{[salt]} are the discontinuous architecture of Sec.~\ref{sec:weak} in use: \pyth{H2} and \pyth{HF} exist only in the salt, and never acquire degrees of freedom in the metal.
The five-channel isotopologue set of Sec.~\ref{sec:isotopes} is assembled the same way, with the mixed-isotopologue degeneracy carried in its rate constant rather than by declaring the channel twice.

\subsection{Parameterisation and numerical behaviour}
\label{sec:params}

Every channel needs forwards and backwards rate constants, and the two come from different sources.
Detailed balance, Eq.~(\ref{eq:detailed_balance_general}), fixes the \emph{ratio} $k_r^{+}/k_r^{-}$ from the same thermodynamic data that parameterise the LTE conditions of Sec.~\ref{sec:lte}, so no new information is needed there.
The \emph{magnitude} is the additional input the framework requires.
It can come from atomistic modelling, from fitting to permeation breakthrough transients, or, as a first estimate for the recombination channel, from the gas/metal recombination coefficients measured for the same metal.
Silva-Sol\'is et al.\ \cite{silva-solis_hydrogen_2026} follow this methodology for W/Cu, fixing the ratios from the solution energies of the interfacial and bulk sites \cite{silva-solis_solution_2024} and fitting only the magnitude to the relaxation time of a site network computed from density functional theory. 
The sensitivity of a prediction to that magnitude is exactly what the Damk\"ohler number of Eq.~(\ref{eq:damkohler}) measures: where $\Da \gg 1$ an order of magnitude in $k^{+}$ is immaterial, and where it is not, the interface is rate-limiting and the number has to be justified.

Two properties of the present implementation are limitations of the code and not of the framework.
The rate constants are prescribed as plain numbers, with no Arrhenius form, so a temperature dependence has to be imposed by the user and the activity $a_\mathrm{F}$ of Eq.~(\ref{eq:model3_rate}) is folded into $k_\mathrm{f}^{+}$ by hand.
More importantly, the ratio $k_r^{+}/k_r^{-}$ is not checked against Eq.~(\ref{eq:detailed_balance_general}) by the solver: thermodynamic consistency is a property of how the model is parameterised, and it remains the responsibility of whoever writes the input.

\section{Verification}
\label{sec:verification}

To verify the implementation of the interface conditions of Sec.~\ref{sec:models}, a simple case is considered: two one-dimensional slabs of equal thickness with a total length, $L = 1$, sharing an interface at the midpoint.
The diffusivities of the two sides are $D_A = 0.5$ and $D_B = 1$ in arbitrary units, and concentrations $c_0 = 2$ and $c_L = 1$ are held at the two outer faces.
Only the chemistry on $\Gamma$ changes between the tests below.

This case has an analytical steady state.
With no source, each slab carries a linear profile, so the interfacial concentrations follow from the flux through the bulk resistance of each side.
Substituting them into the interface condition leaves a single algebraic equation for the flux, linear for Model~1 and quadratic for Model~2 (see \ref{app:analytical}).
P1 elements represent a linear profile exactly, so the bulk discretisation contributes no error; therefore, the differences below are those of the interface term alone, up to the tolerance of the nonlinear solve.

Section~\ref{sec:lte_limit} sweeps the Damk\"ohler number with one channel on the interface, and Sec.~\ref{sec:exponent_verification} opens two channels and sweeps the branching ratio between them.

\subsection{Recovery of LTE in the fast-kinetics limit}
\label{sec:lte_limit}

\begin{figure*}[htbp]
    \centering
    \includegraphics[width=0.8\linewidth]{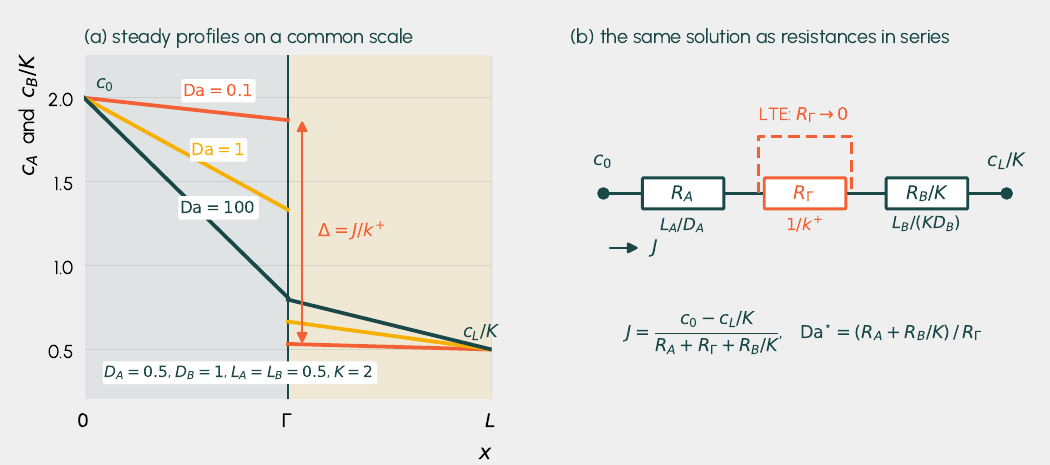}
    \caption{The two-slab verification problem of Sec.~\ref{sec:lte_limit}.
    (a) Steady profiles at three Damk\"ohler numbers, drawn on a common scale so
    that LTE is continuity: $c_A$ on side $A$, $c_B/K$ on side $B$. The jump
    remaining at $\Gamma$ is the defect $\Delta$ of
    Eq.~(\ref{eq:defect_1_explicit}), marked here for the slowest channel.
    (b) The analytical solution Eq.~(\ref{eq:model1_analytical}) reads as three
    resistances in series, LTE being the short circuit of the middle element.}
    \label{fig:two_slab}
\end{figure*}

The first test puts Model 1 on the interface, with $k^{-}/k^{+} = 0.5$.
The steady problem is solved over a range of Damk\"ohler numbers, $\Da = k^{+}L_A/D_A$, from $10^{-2}$ to $10^{6}$.
Plotting the two slabs on a common scale, $c_A$ on one side and $c_B/K$ on the other, turns Eq.~(\ref{eq:lte_ss}) into continuity, so what is left at $\Gamma$ is the departure from LTE $\Delta = J/k^{+}$ (see Fig.~\ref{fig:two_slab}).

\begin{figure}[ht]
    \centering
    \includegraphics[width=1\linewidth]{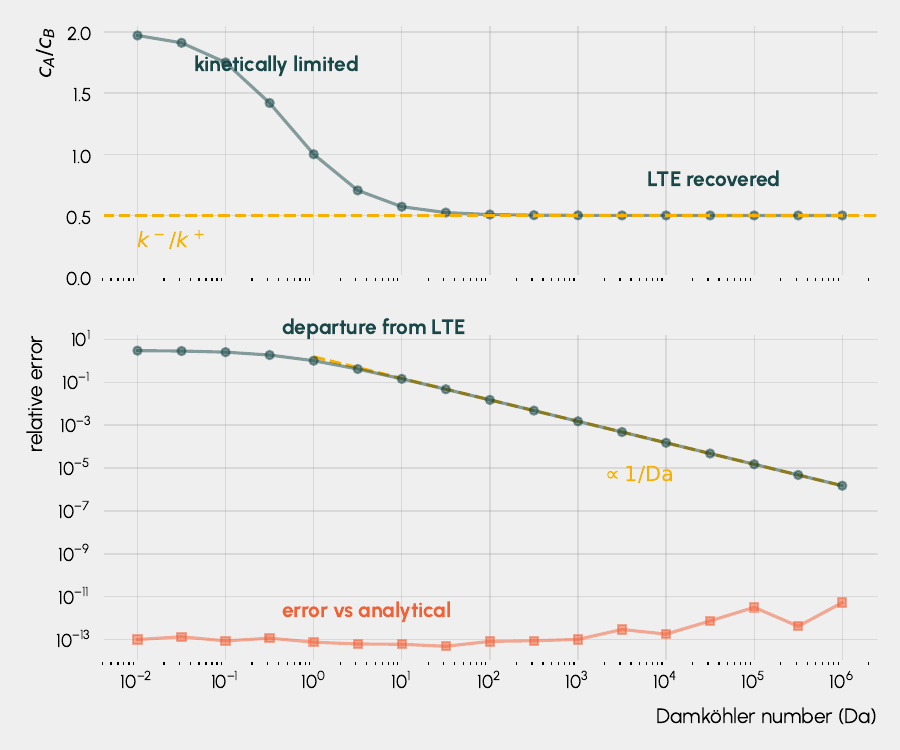}
    \caption{Steady state of the Model 1 problem against $\Da$. Top: the
    interfacial ratio $c_A/c_B$ approaching the LTE value $k^{-}/k^{+}$.
    Bottom: circles, the relative departure of that ratio from the LTE value,
    which decays as $1/\Da$ (dashed); squares, the relative difference between
    the computed interfacial concentrations and the analytical solution of
    Eq.~(\ref{eq:model1_analytical}), at round-off throughout. Model 2 behaves
    the same way on its own equilibrated quantity, as
    Table~\ref{tab:convergence} records.}
    \label{fig:lte_limit}
\end{figure}

\begin{figure}[ht]
    \centering
    \includegraphics[width=1\linewidth]{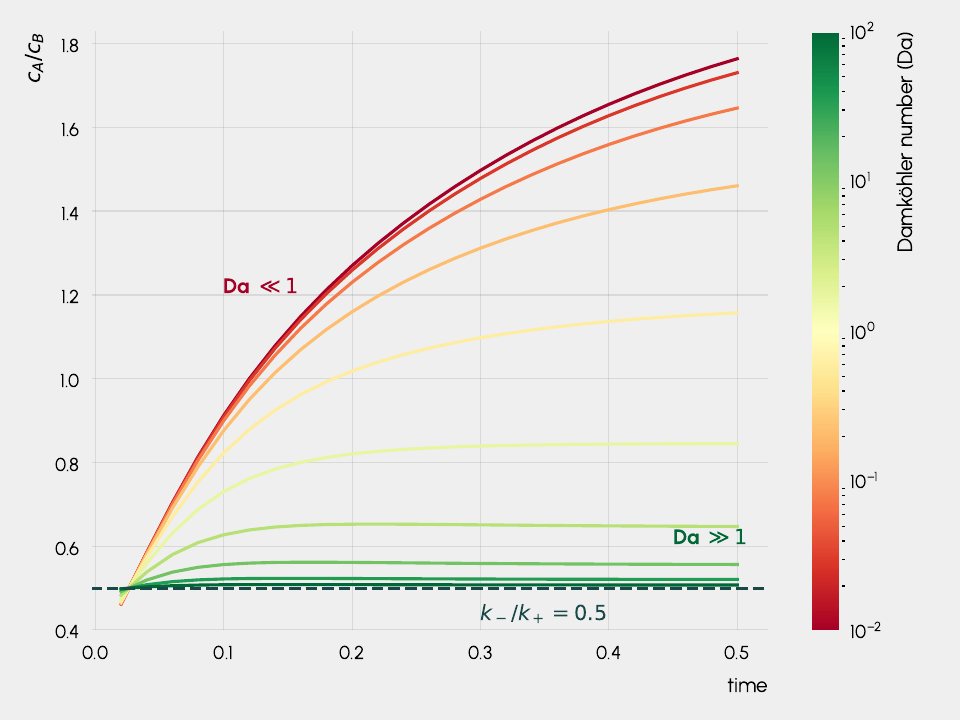}
    \caption{Interfacial ratio $c_A/c_B$ against time on the two-slab problem,
    for Damk\"ohler numbers from $10^{-2}$ to $10^{2}$. The LTE condition
    Eq.~(\ref{eq:lte_ss}) is the dashed line at $k^{-}/k^{+}$. At small $\Da$
    the interface has a relaxation time of its own, and the ratio spends the
    transient far from its equilibrium value; only at large $\Da$ does it sit on
    the LTE line throughout.}
    \label{fig:lte_transient}
\end{figure}

The computed interfacial concentrations agree with the analytical solution to a relative difference of \num{6e-12} or less across the whole sweep.
The departure from LTE, the relative difference between the interfacial ratio and $k^{-}/k^{+}$, decreases as $1/\Da$: from \num{1.5e-2} at $\Da = 10^{2}$ to \num{1.5e-6} at $\Da = 10^{6}$, approaching the predicted asymptote $1.5/\Da$ (see Fig.~\ref{fig:lte_limit} and Table~\ref{tab:convergence}).
It is nonzero at any finite rate constant because the interface carries a flux.
The same sweep run as a transient illustrates the first of the three consequences listed in Sec.~\ref{sec:lte}. 
At moderate $\Da$, the interfacial ratio is far from its equilibrium value for most of the transient, so the interface has a characteristic time of its own, which no algebraic condition can express (see Fig.~\ref{fig:lte_transient}).

The second test replaces the channel with the recombination channel of Eq.~(\ref{eq:model2_rate}) for a metal/liquid pair.
Atomic H lives in the metal slab, [0, 0.5], and the molecular carrier $\mathrm{H_2}$ in the liquid slab, [0.5, 1], so the two sides now hold distinct species.
Eq.~(\ref{eq:detailed_balance_2}) fixes the reverse constant from the forward one, and $K_H/K_S^{2}$ is taken as $0.5$.
Since the channel is second order, $\Da$ is not an input to the problem: we follow Sec.~\ref{sec:damkohler} and sweep the control parameter $2k_\mathrm{r}^{+}c_0L_\mathrm{A}/D_\mathrm{A}$ over the same eight orders of magnitude, reporting the value attained at the interface as a diagnostic.

The computed interfacial concentrations agree with the analytical solution to a relative difference of \num{2.2e-11} or less throughout, and the departure from LTE again falls as $1/\Da$, from \num{4.4e-3} at $\Da = 10^{2}$ to \num{4.4e-7} at $\Da = 10^{6}$, approaching $0.444/\Da$.
Equation~(\ref{eq:lte_sh}) fixes $\cs{\mathrm{H_2}}|_\Gamma/(\cm{\mathrm{H}}|_\Gamma)^{2}$, so that combination is what the departure is measured on here, rather than a ratio of concentrations.
Both channels therefore converge to LTE at the rate Eq.~(\ref{eq:model1_convergence}) predicts, and LTE is recovered continuously and never imposed.

\begin{table}[htbp]
    \centering
    \caption{Convergence to LTE for the two single-channel models on the two-slab problem, over a sweep of eight orders of magnitude in $\Da$.
    The departure is measured on the quantity each channel equilibrates: the ratio $c_A/c_B$ against Eq.~(\ref{eq:lte_ss}) for Model 1, and the Sieverts/Henry combination $\cs{\mathrm{H_2}}|_\Gamma/(\cm{\mathrm{H}}|_\Gamma)^{2}$ against Eq.~(\ref{eq:lte_sh}) for Model 2.
    Both converge first order in $1/\Da$, at the asymptote \ref{app:limits} predicts for these parameters.}
    \label{tab:convergence}
    \begin{tabular}{lll}
         & Model 1 & Model 2 \\
        \hline
        Analytical solution & Eq.~(\ref{eq:model1_analytical})
                    & Eq.~(\ref{eq:model2_analytical}) \\
        Relative difference & $\le \num{6e-12}$ & $\le \num{2.2e-11}$ \\
        Departure at $\Da = 10^{2}$ & \num{1.5e-2} & \num{4.4e-3} \\
        Departure at $\Da = 10^{6}$ & \num{1.5e-6} & \num{4.4e-7} \\
        Order at $\Da = 10^{2}$ & $0.997$ & $0.994$ \\
        Order above $\Da = 10^{4}$ & $1.0000$ & $\ge 0.9999$ \\
        Predicted asymptote & $1.5/\Da$ & $0.444/\Da$ \\
        \hline
    \end{tabular}
\end{table}


\subsection{The apparent exponent of competing channels}
\label{sec:exponent_verification}

The two tests above each equilibrate a single channel.
The third opens two channels on the interface, for which no algebraic interface law exists, and checks the exponents they produce against Eq.~(\ref{eq:n_of_B}).
The liquid now carries two species, $\mathrm{H_2}$ produced by the recombination channel of Eq.~(\ref{eq:model2_rate}) and HF produced by the fluorination channel of Eq.~(\ref{eq:model3_rate}), both declared on the same interface, as in Sec.~\ref{sec:weak}.
Both carriers are held at zero on the outer liquid face, the swept-salt regime in which Eq.~(\ref{eq:n_of_B}) was derived, and they are given deliberately different diffusivities, $D_\mathrm{HF} = D_\mathrm{H_2}/4$, so that the flux and the salt inventory report different exponents and the two can be compared.
The upstream concentration is swept over five orders of magnitude at three values of $a_\mathrm{F}$, taking the branching ratio of Eq.~(\ref{eq:branching}) from \num{2.1e-3} to \num{6.0e3}.

Nothing about the exponent is imposed on the solver, and this is what makes the test meaningful. 
It is measured afterwards as the logarithmic slope of the total atomic flux against the interfacial concentration. 
This is the same kind of slope as that of a log-log plot of measured flux against upstream pressure from a gas-driven permeation experiment.
The computed interfacial concentrations agree with the analytical solution to a relative difference of \num{2.2e-16}, which is machine precision.
The worst case is \num{1.8e-8} at the two smallest concentrations of the most oxidising sweep, where the interfacial concentration is itself \num{4e-4}, and the solver tolerance sets the floor for the absolute error.

\begin{figure}[htbp]
    \centering
    \begin{subfigure}{\linewidth}
        \centering
        \includegraphics[width=\linewidth]{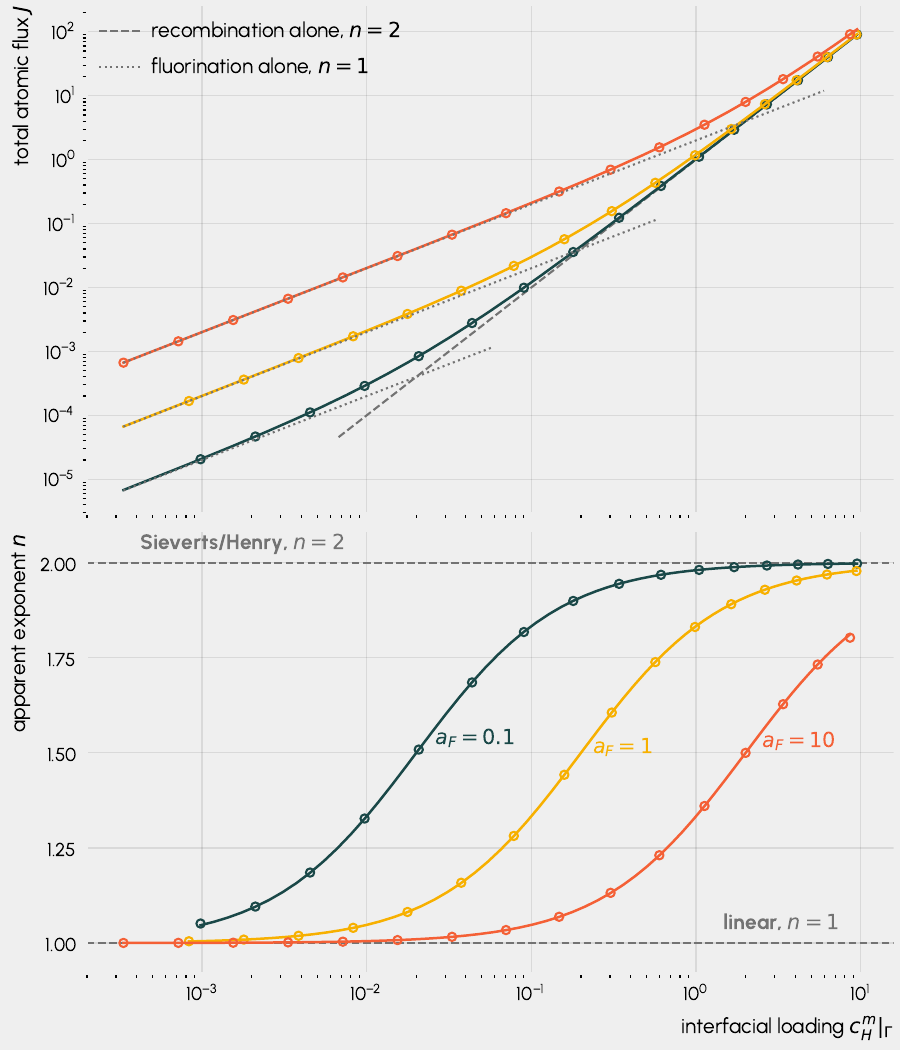}
        \caption{The total atomic flux $J$ leaving the metal against the
        interfacial loading (upper), and the local slope of those curves, the
        apparent exponent $n$ of Eq.~(\ref{eq:apparent_exponent}), on the same
        loading axis and in the same colours (lower), for three values of the
        effective fluoride activity $a_\mathrm{F}$. Lines are the analytical
        solution of \ref{app:analytical} and Eq.~(\ref{eq:n_of_B}), circles the
        values computed by \FESTIM. Each sweep runs along the linear branch of
        its fluorination channel at low loading and joins the quadratic branch
        of the recombination channel at high loading, the same branch for all
        three because $a_\mathrm{F}$ does not enter that channel. The redox
        state sets where the crossover falls, not whether it happens.}
        \label{fig:model3_exponent}
    \end{subfigure}

    \begin{subfigure}{\linewidth}
        \centering
        \includegraphics[width=\linewidth]{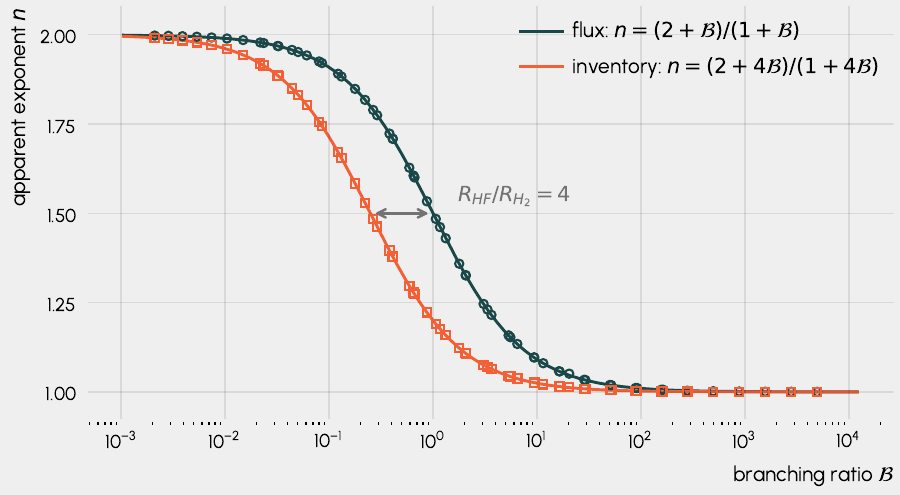}
        \caption{The two ways of reading the exponent, against the branching
        ratio itself. Both follow Eq.~(\ref{eq:n_of_B}), the flux in $\Bra$ and
        the salt-side inventory in $\Bra\,R_\mathrm{HF}/R_\mathrm{H_2}$, so the
        curves have the same shape a constant factor apart: they agree on the
        law and not on the number. The arrow marks the separation of their
        $n = 3/2$ crossings, the resistance ratio, here $4$.}
        \label{fig:model3_branching}
    \end{subfigure}
    \caption{Two channels on one interface, and the exponent they produce.}
    \label{fig:model3}
\end{figure}

The measured exponent tracks the predicted value (see Eq.~(\ref{eq:n_of_B})) to within \num{5.8e-4} over the swept range of branching ratios.
The exponent itself spans from 1 to 2, covering the full range the prediction allows.
At $\Bra = 1$ the measured exponent is $n = 1.5$, the midpoint at which the two channels carry equal atomic fluxes.
The exponent is a continuous function of the branching ratio over its whole range, so a redox sweep moves it, and no fixed choice of $1$ or $2$ describes the interface across such a sweep (see Fig.~\ref{fig:model3_exponent}).
The reverse terms of both channels are active throughout and leave the exponent untouched, because eliminating the salt-side concentrations renormalises each forward constant by a factor independent of the interfacial concentration (\ref{app:analytical}), so the prediction holds beyond the irreversible regime in which it was derived.

An exponent fitted to a sampled salt inventory is not the same number as one measured from the flux.
The interfacial inventory $2\cs{\mathrm{H_2}} + \cs{\mathrm{HF}}$ weights each channel by the downstream resistance its carrier meets, so its logarithmic slope follows the same expression with $\Bra$ replaced by $\Bra\,R_\mathrm{HF}/R_\mathrm{H_2}$, and the sweep reproduces this to \num{5.7e-4}.
Plotted against the same branching ratio, the two curves have identical shape and are separated by the resistance ratio.
The flux exponent passes through $3/2$ at $\Bra = 1$, and the inventory exponent does so at $\Bra = 1/4$ (see Fig.~\ref{fig:model3_branching}).
At a branching ratio of $0.31$, for instance, the flux reports $n = 1.77$ and the inventory $n = 1.45$ on the same interface at the same instant.
The two therefore imply different branching ratios unless the carriers are transported alike, and reading one as though it were the other misplaces $\Bra$ by the resistance ratio.

\section{Application to a representative Ni/FLiBe system}
\label{sec:hyperion}

The tests of Sec.~\ref{sec:verification} are dimensionless.
This section places the framework at a dimensional operating point taken from an experiment, so the groups previously treated as parameters become rate constants at a temperature and geometry that could be tested.

The scope is deliberately narrow. 
Nothing below is fitted, and nothing is compared with a measurement.
\HYPERION{} supplies a realistic geometry, operating conditions and a set of literature transport properties, and the simulations report what the kinetic framework predicts there.
Every result is read against the LTE condition applied to the same problem, and not against data.
Inferring interfacial rate constants from measured transients would require an experimental campaign of its own.

\subsection{System description and model setup}
\label{sec:hyperion_setup}

\HYPERION{} (HYdrogen PERmeatION) at the MIT Plasma Science and Fusion Center \cite{saraswat_permeation_2026} drives hydrogen isotopes through a Ni membrane into a molten FLiBe pool held in a Ni crucible.
The free surface of the salt is swept by a cover gas, and the released flux is obtained from the H concentration in that gas with gas chromatography.

Three interfaces appear in the problem.
The gas/Ni interface uses the existing dissociation-recombination boundary condition, and the FLiBe/cover gas interface releases the carriers from the free surface.
The Ni/FLiBe interface is the one of interest and is the configuration of Sec.~\ref{sec:model3}, with competing recombination and fluorination channels on the same surface.
Calderoni et al.~\cite{calderoni_measurement_2008} measured tritium permeation through Ni into FLiBe and could not determine whether the tritium travelled as a fluoride-bound species, as HT, or as a mixture of the two.
Under LTE, that question has to be answered before the simulation is set up, since the carrier is what the solubility law refers to.
In the kinetic framework, it is a branching ratio, which the model reports.

\begin{figure*}[ht]
    \centering
    \includegraphics[width=1\linewidth]{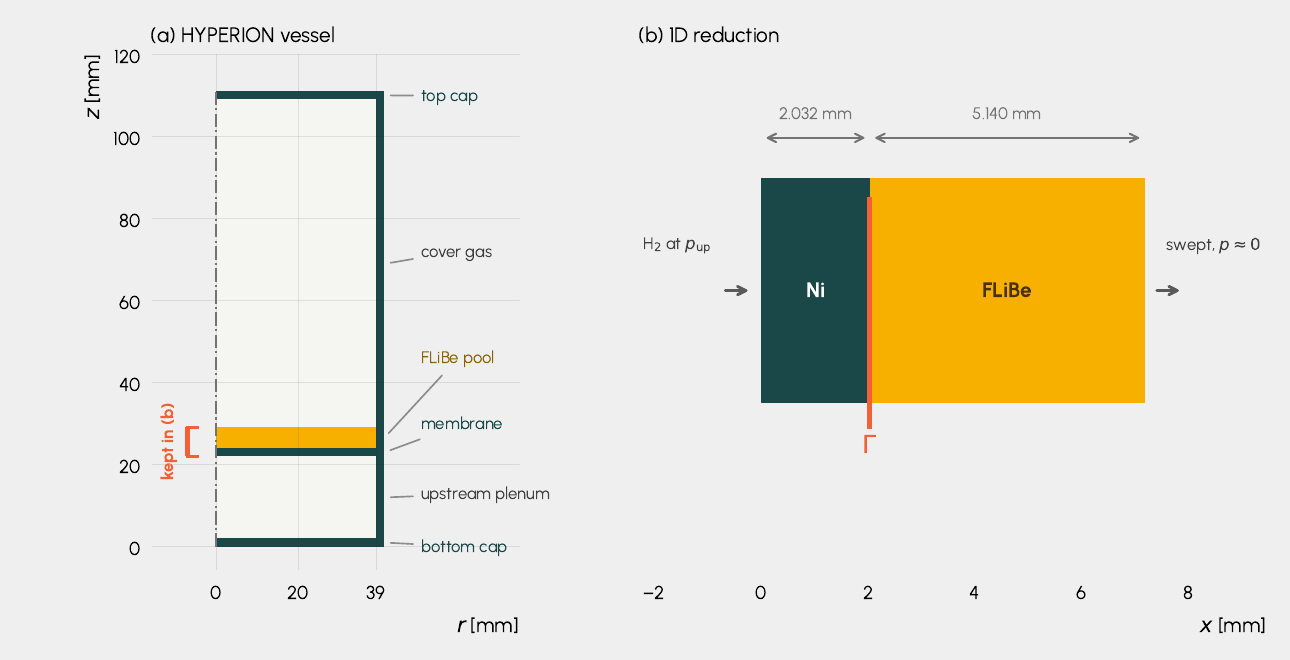}
    \caption{(a) The vessel as an axisymmetric half-section, drawn to the coordinates of the \HYPERION{} modelling repository, with the nickel container, the upstream plenum, the membrane, the FLiBe pool and the cover gas.
    (b) The one-dimensional reduction used in     Sec.~\ref{sec:hyperion}, which keeps the membrane and the pool. The two areas quoted in the figure size the sidewall path that the reduction drops.}
    \label{fig:hyperion_geometry}
\end{figure*}

The vessel is an axisymmetric nickel holding a  FLiBe pool \qty{5.14}{\milli\metre} deep at \qty{500}{\degreeCelsius}, with the membrane below it and a cover-gas space above (Fig.~\ref{fig:hyperion_geometry}).
The one-dimensional model keeps the membrane and the pool and drops the rest.
The multidimensional treatment of that geometry is the subject of Ref.~\cite{yang_quantifying_2026}.

Geometry and operating conditions are taken from the \HYPERION{} reference and modelling repository, and the transport properties are Arrhenius fits from the literature, one source per material (see Table~\ref{tab:hyperion_params}).
The rate constants of the present implementation carry no Arrhenius form (Sec.~\ref{sec:params}), so this whole section is run at a single temperature, \qty{773.15}{\kelvin}, at the low end of the \HYPERION{} window.

\begin{table*}[ht]
    \centering
    \caption{Operating point for the Ni/FLiBe calculations of
    Sec.~\ref{sec:hyperion}, at $T = \qty{773.15}{\kelvin}$. The last column
    records the
    carrier that had to be assumed before each solubility could be reduced to a
    number. That is the point of Sec.~\ref{sec:units}: no solubility row of this
    table is a law-independent property. The last two rows are the same
    measurement counted two ways, a Henry constant counting hydrogen atoms and
    one counting $\mathrm{H_2}$ molecules differing by the stoichiometric factor
    two; the LTE baseline uses the atomic form and Model 2 the molecular one, and
    choosing the wrong one moves the fast-kinetics limit by a factor of two.
    Nothing here is fitted.}
    \label{tab:hyperion_params}
    \begin{tabular}{llll}
        Quantity & Value & Assumed carrier & Source \\
        \hline
        Membrane thickness $L_\mathrm{Ni}$ & \qty{2.032}{\milli\metre} & &
            \HYPERION{} geometry \\
        Salt thickness $L_\mathrm{s}$ & \qty{5.140}{\milli\metre} & &
            \HYPERION{} geometry, \qty{500}{\degreeCelsius} \\
        Membrane radius & \qty{38.99}{\milli\metre} & & \HYPERION{} geometry \\
        Upstream pressure $P_\mathrm{up}$ & \qty{1.31e5}{\pascal} & &
            \HYPERION{} operating range \\
        Downstream & swept, $p \simeq 0$ & & idealisation \\
        $D_\mathrm{Ni}$ & \qty{1.501e-9}{\square\metre\per\second} &
            atomic H & Louthan et al.~\cite{louthan_hydrogen_1975}\\
        $K_{S,\mathrm{Ni}}$ & \qty{2.836e22}{\per\cubic\metre\pascal\tothe{-1/2}} &
            atomic H, Sieverts & Louthan et al.~\cite{louthan_hydrogen_1975} \\
        $D_\mathrm{FLiBe}$ & \qty{1.352e-9}{\square\metre\per\second} &
            unspecified & Calderoni et al.~\cite{calderoni_measurement_2008} \\
        $K_{H,\mathrm{FLiBe}}$ & \qty{2.055e20}{\per\cubic\metre\per\pascal} &
            H atoms, Henry & Calderoni et al.~\cite{calderoni_measurement_2008} \\
        $K_{H,\mathrm{FLiBe}}$ & \qty{1.027e20}{\per\cubic\metre\per\pascal} &
            $\mathrm{H_2}$ molecules, Henry &
            same measurement, recounted \\
        $k_\mathrm{r}^{+}$, $k_\mathrm{f}^{+}a_\mathrm{F}$ & swept &
            see text & Sec.~\ref{sec:hyperion_lte}, \ref{sec:hyperion_branching} \\
        $k_\mathrm{r}^{-}$ & detailed balance &  &
            Eq.~(\ref{eq:detailed_balance_2}) \\
        \hline
    \end{tabular}
\end{table*}

The two bulk diffusion times are $L^{2}/D = \qty{2.75e3}{\second}$ in the metal and \qty{1.95e4}{\second} in the salt, making the salt the slow leg by a factor of seven.
Every transient reported below runs for \qty{e5}{\second}, about five salt diffusion times.

Unlike the recombination channel, the fluorination channel cannot be parameterised solely from material properties.
Even though a Henry constant for HF in FLiBe exists: Field and Shaffer sparged HF into $\qty{66}{\percent}$~LiF--$\qty{34}{\percent}$~$\mathrm{BeF_2}$ and measured the dissolved concentration, finding Henry's law behaviour from \qty{500}{\degreeCelsius} to \qty{700}{\degreeCelsius}~\cite{field_solubilities_1967} (the carrier is unambiguous there, since HF is what was introduced), this measurement cannot supply the other half of the channel's equilibrium constant, the fluorine potential of the salt, which is not a property of FLiBe but a state an operator sets and can move.
The channel is therefore defined by the pair $(k_\mathrm{f}^{+}a_\mathrm{F},\, k_\mathrm{f}^{-})$, and only their ratio is constrained.
We give it no solubility and instead parameterise it in terms of $\Da$ and $\Bra$.

The baseline against which everything below is read treats the interface the way a macroscopic code does, with Sieverts on the nickel against Henry in the salt, Eq.~(\ref{eq:lte_sh}), imposed as the algebraic constraint of Sec.~\ref{sec:lte} through the penalty formulation \FESTIM{} already provides.
At the final time, the computed interfacial concentrations and the downstream flux match the analytical solution of the two-slab LTE problem to relative differences of \num{1.1e-6} and \num{3.4e-6}, respectively, and the interface constraint holds to \num{3.6e-7}.

\subsection{Single isotope, single channel: LTE recovered}
\label{sec:hyperion_lte}

We first switch off the fluorination channel, which corresponds to a strongly reducing salt, and let hydrogen cross the interface by recombination alone.
The channel is that of Eq.~(\ref{eq:model2_rate}), and its reverse constant is fixed by Eq.~(\ref{eq:detailed_balance_2}) from the same $K_S$ and $K_H$ that parameterise the baseline.
The forward constant is then the only number left free, and we report it as the Damk\"ohler number of Eq.~(\ref{eq:damkohler}), $\Da = 2k_\mathrm{r}^{+}\,c^{\star}L_\mathrm{Ni}/D_\mathrm{Ni}$, built as everywhere in this paper at the upstream Sieverts equilibrium $c^{\star} = K_S\sqrt{P_\mathrm{up}}$ of Table~\ref{tab:hyperion_params}.
That makes $\Da$ an input of a sweep instead of an output of one.
The concentration the interface actually reaches is lower, \qty{63}{\percent} of $c^{\star}$ in the equilibrated limit here, so the local Damk\"ohler number sits below the control value, as in the two-slab test of Sec.~\ref{sec:lte_limit}.

Sweeping $\Da$ over six orders of magnitude at fixed thermodynamics, geometry and transport, the computed steady state agrees with the analytical solution of the single-channel problem to a relative difference of \num{7.7e-6} at the slow end of the sweep and to better than \num{5e-6} everywhere else.
The steady flux approaches the LTE one from below as $1/\Da$, as in Sec.\ref{sec:lte_limit}.

\begin{figure*}[ht]
    \centering
    \includegraphics[width=\linewidth]{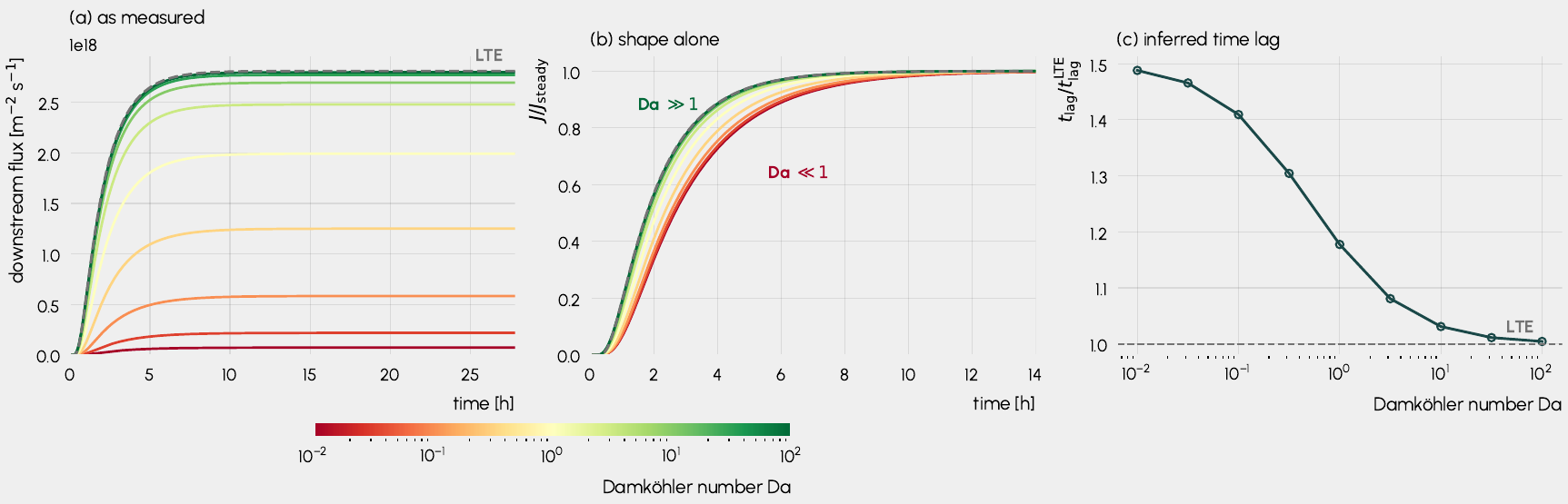}
    \caption{Nine permeation transients at the operating point of
    Table~\ref{tab:hyperion_params}, one recombination channel, $\Da$ swept from
    $10^{-2}$ to $10^{2}$. Every diffusivity, solubility and dimension is held
    fixed across the sweep and detailed balance carries $k_\mathrm{r}^{-}$ along
    with $k_\mathrm{r}^{+}$, so the thermodynamics does not move either.
    (a) The transients as they would be measured, against the LTE reference
    (dashed). (b) The same curves divided by their own steady values, which
    isolates the shape from the amplitude. (c) The time lag of each transient
    relative to the LTE one; since the model diffusivities never change, the
    departure from unity is the interfacial resistance being read as bulk
    transport.}
    \label{fig:hyperion_damkohler}
\end{figure*}

Of the two diagnostics, the steady flux is the more sensitive (Fig.~\ref{fig:hyperion_damkohler}): at $\Da = 10^{-2}$ it is
\qty{2.6}{\percent} of the LTE value, while the time lag is only \qty{49}{\percent} longer than the LTE one.
Applying the classical single-slab inversion $D = L^{2}/6t_\mathrm{lag}$ to each computed transient, as
an experimentalist might do to a measured one, returns an apparent diffusivity \qty{33}{\percent} below the LTE value at $\Da = 10^{-2}$ and \qty{29}{\percent} below it at $\Da = 10^{-1}$.
A time-lag analysis of a permeation curve is therefore not a measure of the bulk transport properties alone, and at this operating point the interface can shift the \textit{inferred} diffusivity by \qty{33}{\percent}, with every diffusivity in the model held fixed.
The discrepancy falls to \qty{3.0}{\percent} at $\Da = 10$ and \qty{0.4}{\percent} at $\Da = 10^{2}$, so the requirement $\Da \gtrsim 100$ read off the regime map of Sec.~\ref{sec:damkohler} is confirmed.

\subsection{Competing channels: the apparent interfacial law}
\label{sec:hyperion_branching}

We now open the fluorination channel of Eq.~(\ref{eq:model3_rate}) alongside the recombination channel.
The bulk is untouched.
Far from equilibrium the branching ratio of Eq.~(\ref{eq:branching}) reduces to $\Bra \rightarrow k_\mathrm{f}^{+}a_\mathrm{F}/(2k_\mathrm{r}^{+} \cm{\mathrm{H}}|_\Gamma)$, so a target $\Bra$ at the reference concentration $c^{\star}$ fixes the forward constant of the F channel through $k_\mathrm{f}^{+}a_\mathrm{F} = 2\Bra\,k_\mathrm{r}^{+}c^{\star}$.
Both groups are built at the same $c^{\star}$, so the Damk\"ohler number of the F channel is not free either but $\Da_\mathrm{F} = \Bra\,\Da$.
The reverse constant needs one further choice.
At steady state, the reverse term enters only through the combination $\rho = k^{-}L_\mathrm{s}/D_\mathrm{s}$, since the swept downstream face gives $c|_\Gamma = wL_\mathrm{s}/D_\mathrm{s}$ for each carrier, and $\rho$ only rescales the forward constant.
We set $\rho_\mathrm{F} = \rho_\mathrm{R}$, placing both channels at the same distance from their own equilibrium.
Both salt-side carriers are given the same diffusivity, for lack of a measurement for HF, so the flux split reported below is set at the interface and not by transport away from it.

\begin{figure*}[htbp]
    \centering
    \includegraphics[width=\linewidth]{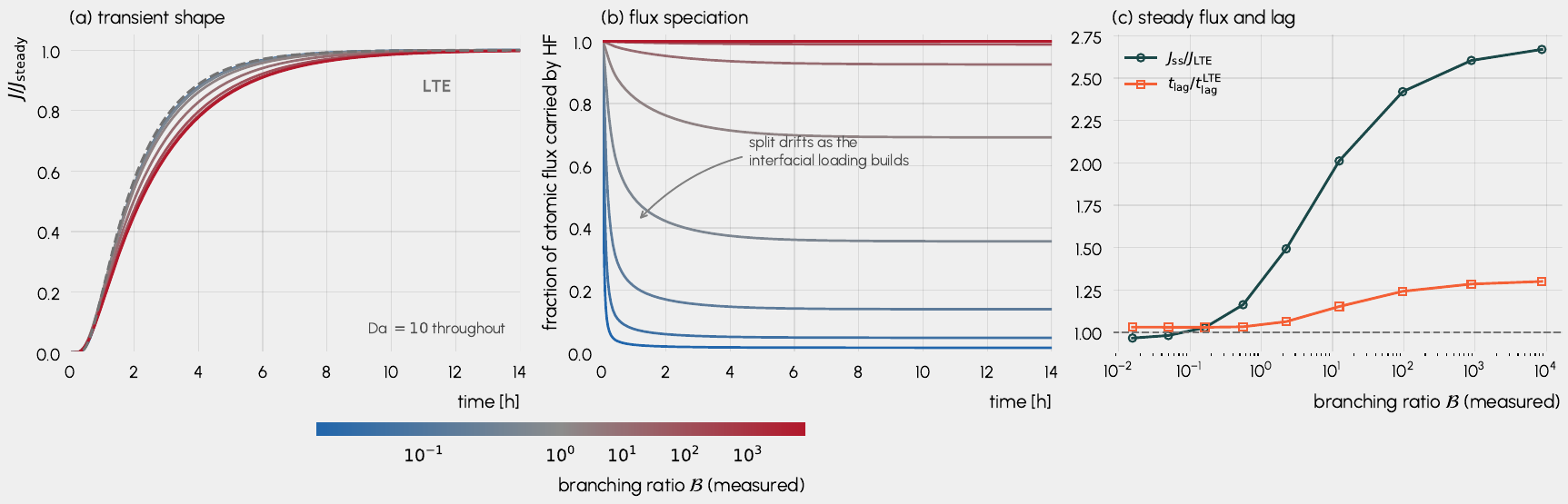}
    \caption{Nine transients with both channels open, $\Da = 10$ fixed on the
    recombination channel and the branching ratio swept over four orders of magnitude. The
    bulk is identical to Fig.~\ref{fig:hyperion_damkohler} and to the LTE
    baseline. (a) Each transient on its own steady value: the redox state alone
    retimes the breakthrough. (b) The fraction of the atomic flux carried by HF
    against time, a direct output of the model that no LTE condition can define.
    (c) The steady flux and the time lag against the measured branching ratio,
    both relative to LTE. Curves are coloured by the measured $\Bra$, blue where
    recombination dominates and red where fluorination does.}
    \label{fig:hyperion_redox}
\end{figure*}

The redox sweep of Fig.~\ref{fig:hyperion_redox} is run at $\Da = 10$, where the recombination channel on its own reaches \qty{96}{\percent} of the LTE flux by Sec.~\ref{sec:hyperion_lte}.
That \qty{4}{\percent} residual is the whole of the slow-channel error at this operating point, and the departures reported below reach a factor of $2.67$, so they are the work of the branching and not of a slow channel.
Three features stand out.

The first is the direction of the error; the steady flux rises \emph{above} the LTE prediction, by a factor $1.49$ at $\Bra \simeq 2.2$ and by $2.67$ at the oxidising end of the sweep.
The fluorination channel is a pathway in parallel with the one the single-carrier model includes, so a Sieverts/Henry interface underestimates the flux by an amount determined by the redox state.
This is testable: raising the fluorine potential at fixed temperature and fixed upstream pressure should raise the permeating flux, which no LTE condition predicts.

The second is that the carrier split drifts through the transient.
Every case starts fluorination-dominated and settles into its steady split because $\Bra$ scales inversely with the interfacial concentration by Eq.~(\ref{eq:branching}), and that concentration builds up as the membrane charges (see Fig.~\ref{fig:hyperion_redox}b).

One consequence of the same drift is that the measured branching ratio departs from the nominal one as soon as the F channel carries a substantial share: a target of $10$ is realised as $96$ and a target of $10^{2}$ as \num{8.6e3}, because the F channel drains the interfacial concentration and $\Bra$ rises as $\cm{\mathrm{H}}|_\Gamma$ falls.
The nominal value is derived from $c^{\star}$, and the true value from the concentration the interface actually reaches.
At low branching, the two differ only by the fixed ratio $c^{\star}/\cm{\mathrm{H}}|_\Gamma \simeq 1.6$ of Sec.~\ref{sec:hyperion_lte}; where the branching matters, they separate by nearly two orders of magnitude.
All results here are reported against the measured value.

\begin{figure*}[htbp]
    \centering
    \includegraphics[width=1\linewidth]{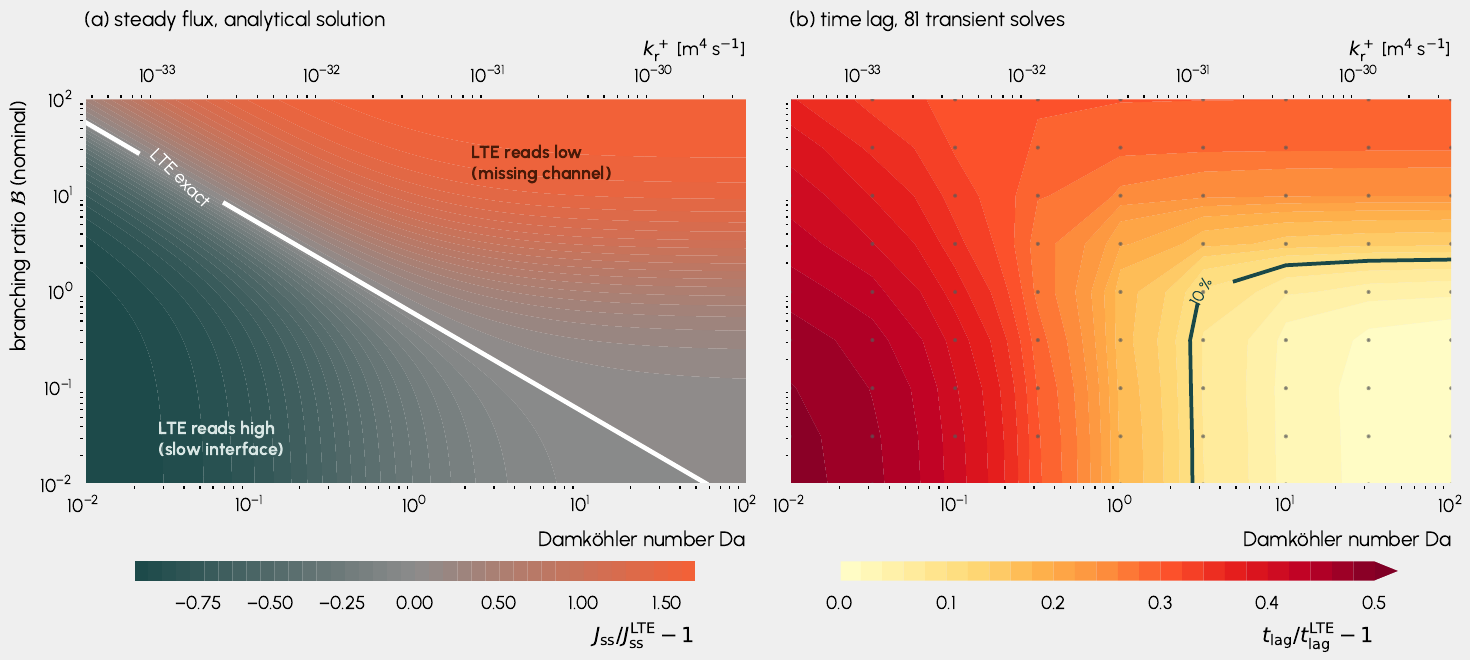}
    \caption{The error of the LTE condition over the $(\Da, \Bra)$ plane at the
    Ni/FLiBe operating point, the dimensional counterpart of
    Fig.~\ref{fig:regime_map}. (a) The signed steady-flux error, from the
    analytical solution; teal where LTE reads high because the interface is slow,
    orange where it reads low because the second channel carries flux LTE has no
    pathway for, and the white contour where the two cancel. (b) The signed
    time-lag error, from 81 transient solves whose sample points are marked. The
    upper axis carries the forward rate constant of the recombination channel
    corresponding to each $\Da$ at this operating point.}
    \label{fig:hyperion_regime_map}
\end{figure*}

Sweeping both groups together yields the map in Fig.~\ref{fig:hyperion_regime_map}, drawn from transient solves of Model 3 on a $9\times9$ grid, with both axes spanning four orders of magnitude.
Two errors are mapped, because an LTE condition can be wrong about the steady state and about the timing independently, and both are signed, since the two failure modes push the steady flux in opposite directions.

Both errors stay below \qty{10}{\percent} at $12$ of the $81$ grid points, the high-$\Da$, low-$\Bra$ corner that LTE assumes it is always in.
At $(\Da, \Bra) = (10^{2}, 10^{2})$ the indicator $\mathcal{E}$ of Eq.~(\ref{eq:lte_indicator}) reports \qty{1}{\percent} against a measured steady-flux error of \qty{167}{\percent}. 
This is why $\mathcal{E}$ is not drawn on the map: its between-channel term is the flux fraction in the weaker channel, so it asks whether \emph{some} single algebraic law could describe the interface and saturates at one half.
The map asks how wrong the \emph{conventional} Sieverts/Henry choice is, and the error of a fixed wrong law does not saturate.

The most striking feature of the map is the diagonal valley in subfigure (a), where the two failure modes cancel.
A slow interface keeps the steady flux below LTE and a second channel lifts it above LTE, and the two are equal and opposite along a line of constant $\Bra\Da$: the crossing sits at $\Bra\Da \simeq 0.5$ across the whole sweep, which by $\Da_\mathrm{F} = \Bra\Da$ is a fixed Damk\"ohler number of the fluorination channel.
Along that line the steady flux is right to a few per cent while the time lag is wrong by up to \qty{38}{\percent} at low $\Da$.
A steady permeation measurement, therefore, cannot validate an LTE condition: agreement on the steady flux is compatible with the interface model being wrong about which species crosses the interface and wrong about when.
The transient is what discriminates, and the valley is the sharpest instance of the distinction drawn in Sec.~\ref{sec:lte} between a steady state and an equilibrium.

\section{Discussion}
\label{sec:discussion}

Sections~\ref{sec:models} to~\ref{sec:hyperion} set out the kinetic framework and apply it to a Ni/FLiBe system.
This section draws out what follows, for the solubility data already in the literature and for deciding when LTE is still valid.

\subsection{A solubility constant has no law-independent units}
\label{sec:units}

The dimensions of a reported solubility follow from the dissolution law assumed in fitting it.
Henry's law, $c = K_H P$, gives $[K_H] = \unit{\mole\per\cubic\metre\per\pascal}$, while Sieverts' law, $c = K_S\sqrt{P}$, gives $[K_S] = \unit{\mole\per\cubic\metre\pascal\tothe{-1/2}}$.
No factor converts one into the other: $K_H = K_S^{2}/c$ holds only at a single pressure.
Any conversion silently fixes a reference pressure and is wrong at every other one.
A tabulated solubility is therefore not a material property. 
It is a property of the material and of the carrier assumed for it, and (A3) is what lets the carrier go unreported.

The most-cited fluoride-melt dataset carries the ambiguity in its own equations.
Calderoni et al.~\cite{calderoni_measurement_2008} reduce their steady-state permeation measurements with their Eq.~(2), $J_{\mathrm{T}} = (D_{\mathrm{FLiBe}} K_{\mathrm{FLiBe}} / L_{\mathrm{FLiBe}}) \sqrt{p_{\mathrm{T_2}}}$, and plot the measured flux against lines of slope one half. 
Dimensional consistency then requires $[K_{\mathrm{FLiBe}}] = \unit{\mole\per\cubic\metre\pascal\tothe{-1/2}}$, a Sieverts constant.
Their Eq.~(7) reports the fitted result as $K_{\mathrm{FLiBe},\mathrm{T}} = \num{7.9e-2}\exp(-\qty{35}{\kilo\joule\per\mole}/R_g T)$ with units of \unit{\mole\per\cubic\metre\per\pascal}, a Henry constant.
Their Fig.~4 plots it on those axes beside deuterium in FLiBe and hydrogen in FLiNaK, both reduced as Henry constants.
The exponent used to extract the number and the exponent implied by the units reported for it differ by one half.

None of this is a criticism of the measurement.
The measured fluxes are not in question, and the square-root reduction is stated plainly in the text.
The problem is with the tabulated constant.
A reader considering only the units recovers a flux linear in pressure.
A reader who takes Eq.~(2) recovers one going as its square root, and the two agree at a single pressure.
Here (A3) has not simply gone unrecorded, since the same paper implies one carrier in its equations and another in its units.

The consequence for fluoride melts is a reported disagreement.
Against the square root Calderoni et al.\ report for FLiBe, Fukada and Morisaki~\cite{satoshi_fukada_hydrogen_2006} measure an almost linear dependence for FLiNaK.
They conclude that hydrogen dissolves and permeates as $\mathrm{H_2}$ because the permeability and the solubility they measure are both almost linear in pressure.
The review literature expects the same, since dissolved diatomic $\mathrm{H_2}$ in a molten salt should obey Henry's law \cite{forsberg_fusion_2020}.
The reasoning is sound under (A2), and it is the assumption our framework removes.
The exponent they measured belongs to the interface at their redox state, and not to FLiNaK.

\subsection{The pressure exponent measures the branching ratio}
\label{sec:exponent_meaning}

A square-root dependence in an ionic melt is not a measurement artefact.
It is the signature of \emph{dissociative} dissolution, that is, of the fluoride-bound carrier Calderoni et al.\ identify~\cite{calderoni_measurement_2008}.
They conclude that tritium does not recombine at the Ni/FLiBe interface at all, and that it moves through the salt in atomic form, bound to $\mathrm{BeF_4^{2-}}$ or as HT. 
They also note that only ionic dissolution accounts for a measured solubility close to HF reference values and far above $\mathrm{H_2}$.
The metal-side loading follows Sieverts' law, $\cm{\mathrm{H}} \propto \sqrt{P}$, so the apparent exponent of Eq.~(\ref{eq:n_of_B}) maps onto the measured pressure dependence.
A fluorination-dominated interface, $\Bra \gg 1$, gives $n \rightarrow 1$ and a square-root response of the permeating flux.
A recombination-dominated one, $\Bra \ll 1$, gives $n \rightarrow 2$ and a flux linear in pressure.
The pressure exponent in a permeation experiment is set by the branching ratio, not by any fixed property of the salt.
The FLiBe and FLiNaK datasets may therefore both be correct and represent different redox states.
This is testable: \emph{a redox sweep at fixed temperature should move the measured pressure exponent continuously between 0.5 and 1} (see Sec.~\ref{sec:exponent_verification}).

A hint of the same effect may already be present within a single dataset.
Analysing Fig.~3 of Ref.~\cite{satoshi_fukada_hydrogen_2006}, the steady permeation flux through FLiNaK at \qty{500}{\celsius}, \qty{600}{\celsius} and \qty{700}{\celsius} over three orders of magnitude of upstream pressure.
Taken over the full range, the logarithmic slope is \num{1.02}, consistent with the linear reading of the original paper.
Between \qty{1}{\kilo\pascal} and \qty{10}{\kilo\pascal} it falls to \num{0.23}, \num{0.54} and \num{0.91}, in that order of temperature.
Recombination is quadratic in $\cm{\mathrm{H}}$ and fluorination is linear, so $\Bra$ grows as the loading falls, and the slope drifts from $1$ towards $0.5$.
Loading increases with both pressure and temperature, so Eq.~(\ref{eq:n_of_B}) predicts the observed ordering.

We propose this only as a possible explanation.
Each temperature has only four points and the figure reports no uncertainties.
Two also fall outside the interval [$0.5$, $1$] that the two channels can produce.
The \num{0.23} at \qty{500}{\celsius} lies below it, and the \num{1.4} and \num{1.7} above \qty{100}{\kilo\pascal}, and account for neither.
Settling this requires the redox sweep proposed above, with more pressures at each temperature and uncertainties reported for each flux.

Similarly, the Ni/FLiBe permeation dataset cannot be ingested into a transport code without first assuming speciation, which we argue is unknown.
We therefore use Ref.~\cite{calderoni_measurement_2008} as qualitative evidence that the interface is not a simple recombination boundary, and not as an independent parameterisation of its kinetics.
Our own rate constants are reported as the dimensionless groups $\Da$ and $\Bra$ for the same reason.
For the community, the recommendation is to report the fitted pressure exponent and the salt redox state alongside any tabulated solubility.
The number then survives a change in the interpretation of speciation.

\subsection{When can LTE still be used?}
\label{sec:when_lte}

The criteria of Sec.~\ref{sec:damkohler} reduce to three questions, to be asked in this order.

\begin{enumerate}
    \item Is there a single exchange channel, or does one dominate (see Fig.~\ref{fig:regime_map})? If not, no single algebraic law describes the interface, however fast the kinetics.
    \item Is each carrier formed from a single mobile species? If not, a per-species LTE is ill-posed.
    \item If both answers are yes, check $\Da \gg 1$ and $J/(k^{+}c) \ll 1$.
\end{enumerate}


\subsection{Limitations and future work}
\label{sec:limitations}

We treat interfaces as carrying no stored inventory.
Eliminating a resolved interfacial population, where it is quasi-steady and few of its sites are filled, recovers the previously used first-order channel, with the ratio $k^{+}/k^{-}$ intact.
The first condition holds exactly at steady state, so only transients depend on it.
What the limit discards is the inventory, which delays breakthrough where it matches the bulk loading, and the site blocking that a finite site density imposes.
This will be the subject of future work, as will the W/Cu~\cite{silva-solis_hydrogen_2026} and Be/BeO~\cite{hodille_molecular_2022} interface-trapping cases.

The activity $a_\mathrm{F}$ is prescribed, with no coupled salt redox or corrosion model behind it.
Since HF production, container corrosion and buffer depletion all influence the activity, two-way coupling with a thermochemical solver is the natural next step.

The demonstration case (Sec.~\ref{sec:hyperion}) is planar and uniform, with no natural convection or gas-bubble effects.
The salt-side transport parameters are not independently measured, since the available FLiBe solubility data are reported under an assumed dissolution law (Sec.~\ref{sec:units}),
What we demonstrate is that the framework is internally consistent, that it reproduces LTE wherever LTE applies, and what it predicts at a representative operating point.
It is not validated against an independent parameterisation: such a parameterisation does not currently exist.

Nothing in the framework is specific to fluoride salts, or to fusion.
The same structure appears at any interface where the transported element changes chemical identity on crossing.

\section{Conclusions}
\label{sec:conclusions}

We have formulated a general kinetic interface framework for hydrogen isotope transport and implemented it in \FESTIM{}.
The algebraic local-equilibrium constraint is replaced by reversible reaction channels at the interface, each obeying mass-action kinetics.
Detailed balance fixes each rate ratio from the same thermodynamic data that parameterise LTE.
LTE is recovered analytically and numerically as the fast-kinetics limit of a \emph{single} channel.
The Sieverts/Sieverts and Sieverts/Henry interface laws currently in use are therefore the limits of two different channels within a single framework, not two models to choose between.

Two dimensionless numbers set when that limit holds: a \emph{Damk\"ohler number} measuring equilibration within a channel, and a \emph{branching ratio} measuring the competition between channels.
A large Damk\"ohler number is necessary but not sufficient.
A permeating interface carries a net flux and is therefore never at equilibrium.
How that flux divides between competing channels is set by the branching ratio, regardless of how fast the kinetics are.
Neither number helps when more than one isotope is present, since a per-species LTE condition is then ill-posed.

At the Ni/FLiBe operating point of \HYPERION{}, hydrogen leaving the metal partitions kinetically between a molecular and a fluoride carrier.
The apparent interfacial law is neither Sieverts nor Henry but drifts between them with loading and with salt redox state.
No fixed LTE condition reproduces that behaviour across a redox sweep.
Because the fluorination channel runs in parallel with the one included in a single-carrier model, an LTE interface not only misestimates the steady flux but also underestimates it by an amount determined by the redox state.
The two failure modes cancel along a line of constant $(\Da, \Bra)$, where the steady flux is right while the time lag is not.
A steady permeation measurement, therefore, cannot validate an LTE condition; the transient is what discriminates.

One consequence for the property database follows.
The pressure exponent measured in a salt permeation experiment is set by the interfacial branching ratio, and not by any fixed property of the salt.
This offers a physical reconciliation of the conflicting square-root and linear pressure dependences reported for fluoride melts.
A redox sweep at a fixed temperature should continuously move the measured pressure exponent between 0.5 and 1.



\section{Data availability}

The manuscript source and all of the simulation code behind this work are openly available at \url{https://github.com/festim-dev/interface-models} and archived on Zenodo~\cite{remi_delaporte_mathurin_2026_22112697}.
Every figure in this paper is produced by a script in that repository, and the parametric sweeps write the tabulated data they plot alongside their figures.

\appendix
\section{The LTE limit}
\label{app:limits}

\paragraph{Model 1} Write $K \equiv k^{+}/k^{-}$, which detailed balance fixes at $K_{S,B}/K_{S,A}$ by Eq.~(\ref{eq:detailed_balance_1}).
The flux condition Eq.~(\ref{eq:model1_flux}) can be rearranged as
\begin{equation}
    \Delta \equiv c_A\big|_\Gamma - \frac{c_B\big|_\Gamma}{K}
    = \frac{J}{k^{+}} ,
    \label{eq:defect_1}
\end{equation}
where $\Delta$ measures the departure from the LTE condition Eq.~(\ref{eq:lte_ss}), which is $\Delta = 0$.
Two things follow immediately.
Taking $k^{+} \rightarrow \infty$ at fixed $K$ sends $\Delta \rightarrow 0$ at any finite flux, the LTE limit.
At finite $k^{+}$ the departure is proportional to the flux the interface carries, so an interface that is transmitting is never at equilibrium however fast its kinetics.
That is the steady-state/equilibrium distinction of Sec.~\ref{sec:lte} in one line.
Substituting the steady two-slab solution for $J$ gives the size of the departure (see \ref{app:analytical}),
\begin{equation}
    \begin{gathered}
        \Delta = \frac{c_0 - c_L/K}{1 + \Da^{\star}} , \\[3pt]
        \Da^{\star} \equiv k^{+}\left(\frac{L_A}{D_A}
                                    + \frac{L_B}{K D_B}\right) ,
    \end{gathered}
    \label{eq:defect_1_explicit}
\end{equation}
with $\Da^{\star}$ the ratio of the total bulk resistance to the interfacial resistance $1/k^{+}$, that is, the Damk\"ohler number of Eq.~(\ref{eq:damkohler}) evaluated over both slabs.
The relative error on the flux is
\begin{equation}
    \begin{aligned}
        \frac{J}{J_\mathrm{LTE}}
        &= \frac{\Da^{\star}}{1 + \Da^{\star}} \\[3pt]
        &= 1 - \frac{1}{\Da^{\star}}
           + \mathcal{O}\!\left(\Da^{\star -2}\right) ,
    \end{aligned}
    \label{eq:model1_convergence}
\end{equation}
so convergence to LTE is monotone and first order in $1/\Da^{\star}$.
This is the rate verified numerically in Sec.~\ref{sec:lte_limit}.

\paragraph{Model 2} The same argument applies to a channel of higher order, with the departure measured on the quantity the channel equilibrates, not on the concentration itself.
Rearranging Eq.~(\ref{eq:model2_rate}) and using Eq.~(\ref{eq:detailed_balance_2}),
\begin{equation}
    \left(\cm{\mathrm{H}}\big|_\Gamma\right)^{2}
    - \frac{K_S^{2}}{K_H}\,\cs{\mathrm{H_2}}\big|_\Gamma
    = \frac{w_\mathrm{rec}}{k_\mathrm{r}^{+}}
    = \frac{J}{2\,k_\mathrm{r}^{+}} ,
    \label{eq:defect_2}
\end{equation}
whose vanishing is exactly the Sieverts/Henry condition Eq.~(\ref{eq:lte_sh}).
Letting $k_\mathrm{r}^{+} \rightarrow \infty$ at fixed $k_\mathrm{r}^{+}/k_\mathrm{r}^{-}$ therefore recovers LTE, again with a leading correction linear in the flux and in the reciprocal rate constant.
The difference with Model 1 is that the effective exchange velocity $2k_\mathrm{r}^{+}\cm{\mathrm{H}}|_\Gamma$ now depends on the solution, so $\Da$ is not an input to the problem but a diagnostic of it, as discussed in Sec.~\ref{sec:damkohler}.

\paragraph{Model 3} Let both channels be fast, $k_\mathrm{r}^{+}$ and $k_\mathrm{f}^{+} \rightarrow \infty$ at fixed ratios.
Each channel then imposes its own algebraic relation on the trace values,
\begin{equation}
    \begin{aligned}
        \cs{\mathrm{H_2}}\big|_\Gamma
            &= \frac{K_H}{K_S^{2}}
               \left(\cm{\mathrm{H}}\big|_\Gamma\right)^{2} , \\[3pt]
        \cs{\mathrm{HF}}\big|_\Gamma
            &= \frac{k_\mathrm{f}^{+}}{k_\mathrm{f}^{-}}\,
               a_\mathrm{F}\,\cm{\mathrm{H}}\big|_\Gamma ,
    \end{aligned}
    \label{eq:model3_fast}
\end{equation}
and there are two of them, one per carrier.
No single relation of the form of Eq.~(\ref{eq:lte_ss}) or Eq.~(\ref{eq:lte_sh}) is recovered: the total salt-side hydrogen $2\cs{\mathrm{H_2}} + \cs{\mathrm{HF}}$ is the sum of a quadratic and a linear term in $\cm{\mathrm{H}}$, and its logarithmic slope is neither of the two values LTE offers but the intermediate exponent computed in \ref{app:analytical}.
The partition of the flux is worse still.
Both $w_\mathrm{rec}$ and $w_\mathrm{F}$ remain finite in the limit, since a fast channel may carry any flux while sitting arbitrarily close to its own equilibrium, and their ratio is $\Bra$, which is fixed by the forward rate constants and not by any equilibrium constant.
Fast kinetics thus removes the departure \emph{within} each channel while leaving the split \emph{between} them undetermined by thermodynamics.
That is the assertion of Sec.~\ref{sec:model3}.

Where the interfacial plane is instead resolved as a population of sites with its own rate equation~\cite{hodille_molecular_2022,silva-solis_hydrogen_2026}, eliminating that population under the assumptions that it is quasi-steady on the transport timescale and dilute returns Eq.~(\ref{eq:model1_flux}), with $k^{\pm}$ built from the individual hop rates and Eq.~(\ref{eq:detailed_balance_1}) unchanged, the site energy of the plane cancelling from the ratio and setting only the magnitude.
Model 1 is in that sense the adiabatic elimination of such a description, not an alternative to it.

\section{Analytical steady-state solutions}
\label{app:analytical}

\paragraph{Model 1, two slabs in series} Take $\Omega_A = [0, L_A]$ and $\Omega_B = [L_A, L_A + L_B]$ with the interface at $x = L_A$, concentrations $c_0$ and $c_L$ imposed at the outer faces, and no source.
At steady state the profile in each slab is linear and the same flux $J$ crosses both, so
\begin{equation}
    \begin{aligned}
        c_A\big|_\Gamma &= c_0 - J R_A ,
            &\qquad R_A &\equiv \frac{L_A}{D_A} , \\[3pt]
        c_B\big|_\Gamma &= c_L + J R_B ,
            &\qquad R_B &\equiv \frac{L_B}{D_B} .
    \end{aligned}
    \label{eq:two_slab_traces}
\end{equation}
Substituting into Eq.~(\ref{eq:model1_flux}) and solving for $J$ gives
\begin{equation}
    \begin{aligned}
        J &= \frac{k^{+}c_0 - k^{-}c_L}{1 + k^{+}R_A + k^{-}R_B} \\[3pt]
          &= \frac{c_0 - c_L/K}{R_A + R_B/K + 1/k^{+}} ,
    \end{aligned}
    \label{eq:model1_analytical}
\end{equation}
with $K = k^{+}/k^{-}$ as in \ref{app:limits}.
The second form is the useful one.
The numerator is the difference of the two outer concentrations expressed on a common scale, and the denominator is a sum of three resistances in series: the bulk resistance of each slab, and an interfacial resistance $1/k^{+}$ contributed by the channel.
Setting $1/k^{+} = 0$ recovers the LTE result, so \emph{LTE is the zero-interfacial-resistance limit} of Model 1, and Eqs.~(\ref{eq:defect_1_explicit}) and~(\ref{eq:model1_convergence}) follow by substitution.
The interfacial concentrations are then given by Eq.~(\ref{eq:two_slab_traces}).

\paragraph{Model 2, two slabs in series} Take the same geometry with the metal on $[0, L_\mathrm{m}]$ and the liquid on $[L_\mathrm{m}, L_\mathrm{m} + L_\mathrm{s}]$, $c_0$ imposed on the metal face and $c_L$ on the liquid face, and write $R_\mathrm{m} = L_\mathrm{m}/D^{\mathrm{m}}$ and $R_\mathrm{s} = L_\mathrm{s}/D^{\mathrm{s}}$.
Both profiles are again linear at steady state, but the two slabs now carry different fluxes: the atomic flux $2w_\mathrm{rec}$ crosses the metal and the molecular flux $w_\mathrm{rec}$ crosses the liquid, so
\begin{equation}
    \begin{aligned}
        \cm{\mathrm{H}}\big|_\Gamma &= c_0 - 2w_\mathrm{rec}R_\mathrm{m} , \\[3pt]
        \cs{\mathrm{H_2}}\big|_\Gamma &= c_L + w_\mathrm{rec}R_\mathrm{s} .
    \end{aligned}
    \label{eq:two_slab_traces_2}
\end{equation}
Substituting into Eq.~(\ref{eq:model2_rate}) gives a quadratic in the rate,
\begin{equation}
    \begin{split}
        4k_\mathrm{r}^{+}R_\mathrm{m}^{2}\,w_\mathrm{rec}^{2}
        &- \left(1 + 4k_\mathrm{r}^{+}c_0R_\mathrm{m}
                  + k_\mathrm{r}^{-}R_\mathrm{s}\right)w_\mathrm{rec} \\
        &+ \left(k_\mathrm{r}^{+}c_0^{2} - k_\mathrm{r}^{-}c_L\right) = 0 ,
    \end{split}
    \label{eq:model2_analytical}
\end{equation}
of which the physical solution is the smaller root, the one that remains finite as $R_\mathrm{m} \rightarrow 0$.
Two differences with Model 1 follow.
There is no factorisation of Eq.~(\ref{eq:model2_analytical}) into a driving force over a sum of resistances, so the series reading of Fig.~\ref{fig:two_slab}b holds for Model 2 only after linearisation about a chosen interfacial concentration.
And the LTE limit is not obtained by deleting a term: taking $k_\mathrm{r}^{+},k_\mathrm{r}^{-} \rightarrow \infty$ at fixed $K = k_\mathrm{r}^{+}/k_\mathrm{r}^{-} = K_H/K_S^{2}$ leaves the quadratic
\begin{equation}
    4KR_\mathrm{m}^{2}w_\mathrm{rec}^{2}
    - \left(4Kc_0R_\mathrm{m} + R_\mathrm{s}\right)w_\mathrm{rec}
    + \left(Kc_0^{2} - c_L\right) = 0 ,
    \label{eq:model2_lte_quadratic}
\end{equation}
whose root is the Sieverts/Henry steady state that the sweep of Sec.~\ref{sec:lte_limit} converges to.

\paragraph{Model 3, apparent exponent} Consider the two competing channels of Sec.~\ref{sec:model3} at steady state, in the regime where the reverse terms are small because the salt side is swept, and write $c \equiv \cm{\mathrm{H}}|_\Gamma$.
The atomic flux carried by each channel is
\begin{equation}
    \begin{aligned}
        J_\mathrm{R} &= 2 w_\mathrm{rec} = 2 k_\mathrm{r}^{+} c^{2} , \\[3pt]
        J_\mathrm{F} &= w_\mathrm{F} = k_\mathrm{f}^{+} a_\mathrm{F}\, c ,
    \end{aligned}
    \label{eq:channel_fluxes}
\end{equation}
so that $\Bra = J_\mathrm{F}/J_\mathrm{R}$, in agreement with Eq.~(\ref{eq:branching}).
The total is $J = J_\mathrm{R} + J_\mathrm{F}$, and since $J_\mathrm{R}$ is quadratic in $c$ and $J_\mathrm{F}$ is linear, the apparent exponent of Eq.~(\ref{eq:apparent_exponent}) is a weighted mean of the two orders,
\begin{equation}
    \begin{aligned}
        n = \frac{\partial \ln J}{\partial \ln c}
          &= \frac{2 J_\mathrm{R} + J_\mathrm{F}}{J_\mathrm{R} + J_\mathrm{F}} \\[3pt]
          &= \frac{2 + \Bra}{1 + \Bra} .
    \end{aligned}
    \label{eq:n_of_B}
\end{equation}
The exponent falls monotonically from $n = 2$ at $\Bra \rightarrow 0$, where recombination dominates and the Sieverts/Henry condition Eq.~(\ref{eq:lte_sh}) applies, to $n = 1$ at $\Bra \rightarrow \infty$, where fluorination does and the interface law is linear, passing through $n = 3/2$ at $\Bra = 1$.
Every value in between is attainable, and since $\Bra$ depends on $c$ through Eq.~(\ref{eq:branching}), a single interface visits a range of $n$ as its loading changes.

If the exponent is read off concentrations instead of fluxes, as it is when a solubility is fitted to a sampled salt inventory, the same expression holds with $\Bra$ replaced by $\Bra\,(R_\mathrm{HF}/R_\mathrm{H_2})$, where the $R$ are the downstream transport resistances of the two carriers, since at steady state each salt-side concentration is its production rate times the resistance it sees.
The two agree only when the two carriers are transported alike.

Neither expression requires the reverse terms to be neglected.
Eliminating the salt-side traces, as in Eq.~(\ref{eq:two_slab_traces_2}), makes each rate an explicit function of $c$ alone, $w_\mathrm{rec} = \alpha c^{2} - a_0$ and $w_\mathrm{F} = \beta c - b_0$, with $\alpha = k_\mathrm{r}^{+}/(1 + k_\mathrm{r}^{-}R_\mathrm{H_2})$ and $\beta$ the corresponding renormalisation of $k_\mathrm{f}^{+}a_\mathrm{F}$.
When the salt side is swept the constants $a_0$ and $b_0$ vanish, the two atomic fluxes stay exactly quadratic and exactly linear in $c$, and Eq.~(\ref{eq:n_of_B}) holds with the reverse terms fully active.
The downstream resistance rescales each channel without moving the exponent it contributes.

\bibliographystyle{elsarticle-num}
\bibliography{references}

\end{document}